\documentclass{elsarticle}
\usepackage[utf8]{inputenc}
\usepackage{graphicx}
\usepackage{caption}
\usepackage{subcaption}
\usepackage{amsmath}
\usepackage{bbm}
\usepackage{cleveref}
\usepackage{url}
\usepackage{color}
\usepackage[normalem]{ulem}

\newcommand{\bsym}[1]{\boldsymbol{#1}}

\begin{document}

\title{Machine learning materials physics: Integrable deep neural networks enable scale bridging by learning free energy functions}

\author[umme]{G.H. Teichert}
\author[ucsb]{A.R. Natarajan}
\author[ucsb]{A. Van der Ven}
\author[umme,umm,micde]{K. Garikipati\corref{mycorrespondingauthor}}
\ead{krishna@umich.edu}
\cortext[mycorrespondingauthor]{Corresponding Author}

\address[umme]{Department of Mechanical Engineering, University of Michigan}
\address[umm]{Department of Mathematics, University of Michigan}
\address[ucsb]{Materials Department, University of California, Santa Barbara}
\address[micde]{Michigan Institute for Computational Discovery \& Engineering, University of Michigan}

\begin{abstract}
The free energy of a system is central to many material models. Although free energy data is not generally found directly, its derivatives can be observed or calculated. In this work, we present an Integrable Deep Neural Network (IDNN) that can be trained to derivative data obtained from atomic scale models and statistical mechanics, then analytically integrated to recover an accurate representation of the free energy. The IDNN is demonstrated by training to the chemical potential data of a binary alloy with B2 ordering. The resulting DNN representation of the free energy is used in a mesoscopic, phase field simulation and found to predict the appropriate formation of antiphase boundaries in the material. In contrast, a B-spline representation of the same data failed to resolve the physics of the system with sufficient fidelity to resolve the antiphase boundaries. Since the fine scale physics harbors complexity that emerges through the free energy in coarser-grained descriptions, the IDNN represents a framework for scale bridging in materials systems.
\end{abstract}


\maketitle

\section{Introduction}
An accurate description of the free energy plays a critical role in many physics-based models of materials systems. The Euler-Lagrange equations of stationary problems are obtained by requiring that the first variational derivative of the free energy functional vanish. In the example of elasticity, this leads to the equilibrium equation satisfied by the displacement field \cite{Marsden1994}. Evolution equations can also require first variations of the free energy as inputs. For example, the variational derivatives of the free energy with respect to composition and order parameters define the chemical potential fields used in phase field dynamics \cite{CahnHilliard1958,Allen1979,Provatas2011}. A related result is that the first derivative of the free energy density function with respect to appropriate strain measures gives the conjugate stress. Second derivatives with respect to the temperature yield the heat capacity, those with respect to strains define elasticities \cite{Hillert2007}.

High-fidelity evaluations of free energy density functions are attainable by a combination of atomic scale models (including quantum mechanics and various molecular models) and statistical mechanics. The use of such free energy density functions in the partial differential equations of continuum physics at coarse-grained scales, such as in phase field models, other mass and heat transport equations, and the equations of nonlinear elasticity is a rigorous, if classical approach to scale bridging. In principle, if highly accurate free energy density functions were available, quantitative predictivity would become accessible in a specialized but significant spectrum of mechano-chemically driven phenomena, such as phase transformations, nonlinear and strain gradient elasticity.
    
Implementing these computations, however, can be challenging for a number of reasons. Free energy data is often computed at individual points rather than as an analytic function directly. As such, it becomes necessary to use a fitting process to create a faithful mathematical model of the free energy. The free energy density can be a function of multiple variables, including composition, temperature, strain, and order parameters, thus leading to a high-dimensional function \cite{Rudrarajuetal2016,Sagiyamaetal2016}. Additionally, realistic data for the free energy can contain regions with rapid fluctuations, along with other regions with very gradual slopes \cite{Teichert2017}. Finally, because of the importance of the derivatives of the free energy, it is desirable that the fitting function be smooth. All of these considerations pose challenges to the fitting technique.

Machine learning methods are readily applicable to this problem, and we specifically consider Deep Neural Networks (DNNs) \cite{LeCun2015}. DNNs have been successfully applied to problems with large numbers of inputs, such as RGB pixel values in an image \cite{Krizhevsky2012}, the log-power spectra of speech data \cite{Yong2015}, the configurational energy of multicomponent alloys \cite{natarajan2018}, predicting precipitate morphologies \cite{Teichert2018a} and biomarkers for predicting human age \cite{Putin2016}, among numerous other applications. Because the activation functions used in DNNs are generally global functions with one local feature, DNNs are capable of capturing local phenomena without negatively affecting longer range attributes of the data. Also, with the proper choice of activation functions, DNNs are infinitely differentiable, thus allowing all necessary derivatives to be computed.

One potential disadvantage of DNNs is that they are not, in general, analytically integrable. This is a particular challenge in the case of fitting free energy data, because the free energy is often not directly measured or computed from atomic models and statistical mechanics. Instead, the derivatives of the free energy (i.e. the chemical potentials) are first observed or computed, then integrated to find the free energy of the system \cite{DeHoff2006,Sanchez1984,deFontaine1994,VanderVen1998,vandewalle2002b,VanderVen2010,Puchala2013,Chen2015,Natarajan2016,Teichert2017}. Such an approach is of particular importance in cases where the chemical potential representation must be integrated with respect to chemical variables to obtain the free energy density, whose derivatives with respect to strain then yield the stress for elasticity \cite{Rudrarajuetal2016}. In order to preserve as much information about the derivatives as possible, it is ideal to train directly to the derivative data itself, as opposed to a numerically integrated data set. This requires an alternate form of DNN that can be trained to derivative data, then analytically integrated to represent the free energy itself. We present such an approach in this work.

% It is possible to reduce the complexity required of the DNN by incorporating certain terms that are known beforehand. For example, the ideal solution energy contains logarithmic terms that cause the chemical potentials to diverge at certain locations in the composition-order parameter space. Rather than requiring the DNN to learn this behavior, it can be imposed by incorporating the ideal solution free energy terms themselves into the final expression of the free energy.

In this first presentation of our proposed framework, we consider the problem of chemistry, postponing coupled problems for later communications. To demonstrate the method of training a DNN to derivative data of the free energy, we consider a simple binary substitutional alloy with B2 ordering on a BCC parent structure \cite{Natarajan2017}. The resulting free energy, as a function of composition and an order parameter, is then used in the Cahn-Hilliard and Allen-Cahn phase field equations.

In Section \ref{sec:DNN}, we present the form of an Integrable Deep Neural Network (IDNN). Section \ref{sec:B2} describes the B2 material system and the method of calculating the chemical potential data via a combination of atomic scale and statistical mechanics computations. In this context, the IDNN training to the chemical potential data and the resulting free energy DNN are shown in Section \ref{sec:training}. We describe the phase field formulation in Section \ref{sec:phase field} and present the computational results obtained using the DNN representation of the free energy. Section \ref{sec:splines} describes a process for fitting the chemical potential data using B-splines and compares the resulting fit and phase field simulation with those of the DNN. Concluding discussions are presented in Section \ref{sec:conclusions}.

\section{Integrable Deep Neural Network}
\label{sec:DNN}
\begin{figure}[tb]
    \centering
    \includegraphics[width=0.75\textwidth]{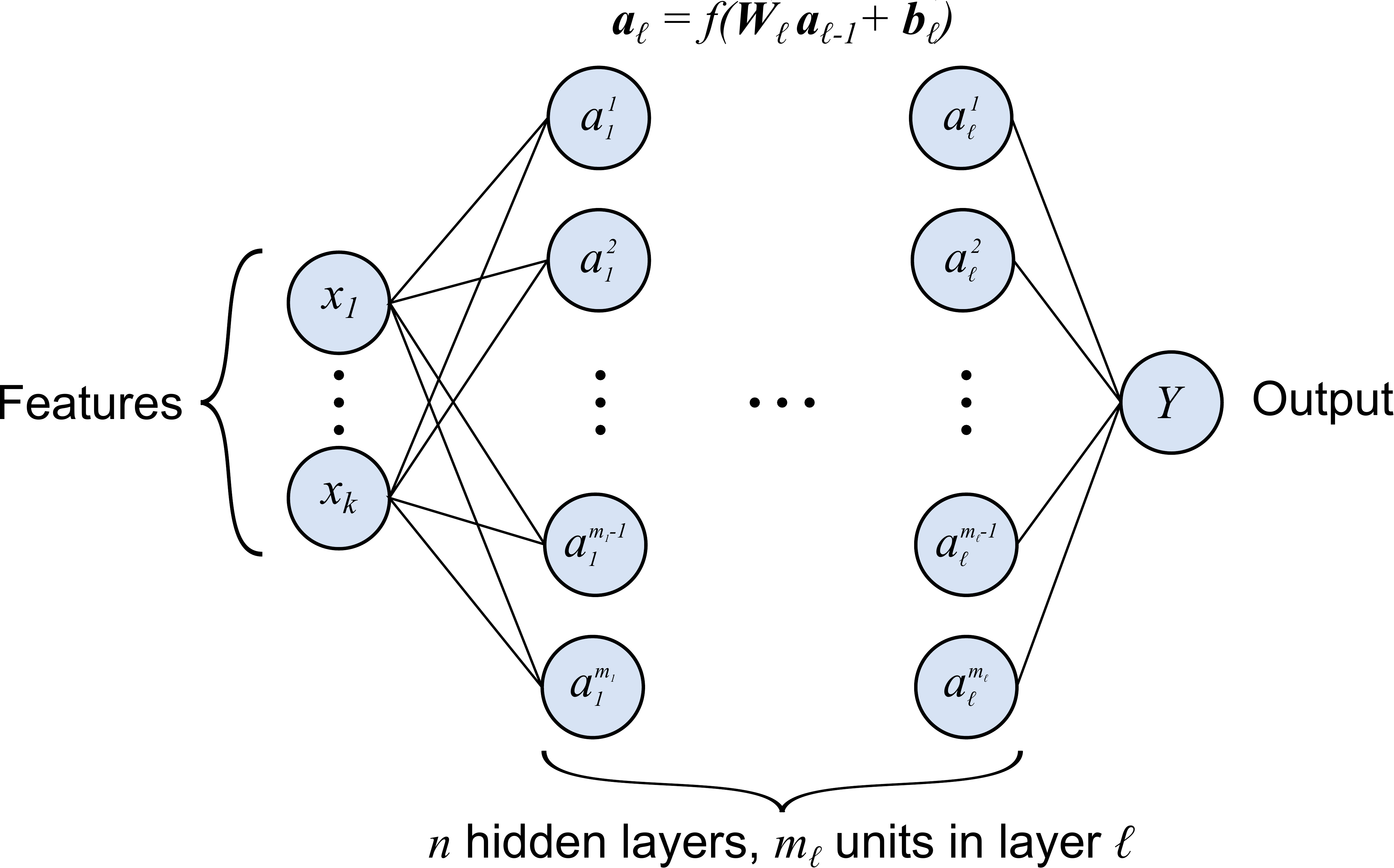}
    \caption{Schematic of a standard, fully connected Deep Neural Network (DNN). In this instance, the output is a scalar.}
  \label{fig:DNN}
\end{figure}

For a fully connected DNN, the following equations are commonly used to define the activation value of unit $i$ in layer $\ell$, denoted here by $a^\ell_i$:
\begin{align}
    a^\ell_i &= f(z^\ell_i)\label{eq:fz}\\
    z^\ell_i &= b^\ell_{i} + \sum_{j=1}^{m_{\ell-1}} W^\ell_{i,j}a^{\ell-1}_{j}\label{eq:z}
\end{align}
where $b^\ell_{i}$ is the bias, $W^\ell_{i,j}$ is the weight, $m_{\ell-1}$ is the number of units in layer $\ell-1$, and $f(\cdot)$ is the activation function (see Figure \ref{fig:DNN}). In a DNN used for regression, the output $Y_i$ is computed using the activation units from the final layer without an activation function:
\begin{align}
    Y_i = b^{n+1}_{i} + \sum_{j=1}^{m_{n}} W^{n+1}_{i,j}a^{n}_{j}
\end{align}

The DNN can be thought of as a function of inputs $\bsym{x}$, weights $\bsym{W}$, and biases $\bsym{b}$, i.e. $\bsym{Y}=\bsym{Y}(\bsym{x},\bsym{W},\bsym{b})$. Training the DNN consists of optimizing the weights and biases to minimize the loss function for a given dataset $\{(\hat{\bsym{x}}_\theta,\hat{\bsym{Y}}_\theta)\}_{\theta=1}^{N_\text{data}}$. The loss function is generally the mean square error (MSE) for regression problems. We can represent this as follows:
\begin{align}
\bsym{\hat{W}},\bsym{\hat{b}} = \underset{\bsym{W},\bsym{b}}{\mathrm{arg\,min}}\,\mathrm{MSE}\left(\bsym{Y}(\bsym{x},\bsym{W},\bsym{b})\Big |_{\bsym{\hat{x}}_\theta},\bsym{\hat{Y}}_\theta\right)
\label{eq:DNN-Wb}
\end{align}
where the sum over $\theta = 1,\dots N_\text{data}$ is implied in the definition of the MSE.

As argued in the Introduction, it can be desirable to train a function with data describing the derivatives. While it is possible to train a standard DNN directly to the derivative data, there are at least two drawbacks. The first is that the standard DNN cannot, in general, be analytically integrated to recover the originating function. Furthermore, in the multidimensional case with multiple sets of partial derivative data, the trained standard DNNs representing the partial derivatives would not necessarily be themselves the derivatives of the same function. This inconsistency poses a problem even in situations where mathematical representations of only the partial derivatives are needed and not the integral itself.

Both of these difficulties can be overcome by differentiating the standard DNN with respect to the desired input variables, say $x_k$ with $k=1,\ldots,n$, then training these differentiated functions to the (partial) derivative data $\{(\hat{\bsym{x}}_\theta,\hat{\bsym{y}}_{k_\theta})\}$, as represented by the following:
\begin{align}
\bsym{\hat{W}},\bsym{\hat{b}} = \underset{\bsym{W},\bsym{b}}{\mathrm{arg\,min}}\,\sum_{k=1}^n\mathrm{MSE}\left(\frac{\partial\bsym{Y}(\bsym{x},\bsym{W},\bsym{b})}{\partial x_k}\Big |_{\bsym{\hat{x}}_\theta},\bsym{\hat{y}}_{k_\theta}\right)
\label{eq:IDNN-Wb}
\end{align}

The functional form that results from differentiating the standard DNN is, of course, analytically integrable. As such, it will be referred to in this work as an IDNN (Integrable Deep Neural Network). The antiderivative of the IDNN is simply a standard DNN with the weights and biases of the trained IDNN. The IDNN has the following form:
\begin{align}
    \frac{\partial a^\ell_i}{\partial x_k} &= f'(z^\ell_i)\frac{\partial z^\ell_i}{\partial x_k} \label{eqn:IDNN1}\\
    %z^\ell_i &= b^\ell_{i} + \sum_{j=1}^{m_{\ell-1}} W^\ell_{i,j}a^{\ell-1}_{j}\\
    a^\ell_i &= f(z^\ell_i)\\
    \frac{\partial z^\ell_i}{\partial x_k} &= \sum_{j=1}^{m_{\ell-1}} W^\ell_{i,j}\frac{\partial a^{\ell-1}_{j}}{\partial x_k} \label{eqn:IDNN4}
\end{align}
where $z^\ell_i$ is given by Equation \eqref{eq:z}. In a slightly more abstracted form with two sets of activation units, $\alpha$ and $\beta_k$, we have:
\begin{align}
    \beta^\ell_{k_i} &= f'(z^\ell_i)\sum_{j=1}^{m_{\ell-1}} W^\ell_{i,j}\beta^{\ell-1}_{k_j}\\
    %z^\ell_i &= b^\ell_{i} + \sum_{j=1}^{m_{\ell-1}} W^\ell_{i,j}\alpha^{\ell-1}_{j}\\
    \alpha^\ell_i &= f(z^\ell_i)
\end{align}
The values of the trained derivative function $y_{i,k} := \partial Y_i/\partial x_k$ and its integral (within an integration constant) $Y_i$ are found as follows:
\begin{align}
    y_{i,k} &= \sum_{j=1}^{m_{n}} W^{n+1}_{i,j}\beta^{n}_{k_j}\\
    Y_i &= \sum_{j=1}^{m_{n}} W^{n+1}_{i,j}a^{n}_{j}
\end{align}
Note that the following are used to compute the activation values for the first hidden layer:
\begin{align}
    \beta^1_{k_i} &= f'(z^1_i) W^1_{i,k}\\
    z^1_i &= b^1_{i} + \sum_{j=1}^{m_{1}} W^1_{i,j}x_{j}\\
    \alpha^1_i &= f(z^1_i)
\end{align}
The activation function, $f(\cdot)$, can be chosen such that its derivative, $f'(\cdot)$, is also a common activation function. For example, with the SoftPlus activation function, $f(x) := \ln(1 + e^x)$, the derivative, $f'(x)$ is the commonly used logistic function, i.e. $f'(x) = 1/(1 + e^{-x})$.

We emphasize that the form of the IDNN in Eq.\,(\ref{eqn:IDNN1})--(\ref{eqn:IDNN4}) has been chosen such that its integral has the form of a standard DNN, where both the IDNN and its integral use the same weights and biases, as is clear from the optimizations \eqref{eq:IDNN-Wb} and \eqref{eq:DNN-Wb}. Thus, once the weights and biases of an IDNN have been trained using \eqref{eq:IDNN-Wb}, the analytically integrated DNN is constructed by simply using the IDNN's weights and biases in a standard DNN given by \eqref{eq:fz} and \eqref{eq:z}.

\subsection{Enforcing symmetries}
%Sometimes it is desirable to enforce certain symmetries on the DNN by first transforming the inputs by a particular function. For example, if $Y$ should be symmetric about $x_1$, one could use $h(x_1) := x_1^2$ as a DNN feature instead of $x_1$ itself. Since the given derivative data is probably differentiated with respect to, for example, $x_1$ and not $h(x_1)$, we must incorporate $h(\cdot)$ into the loss function, e.g. $\mathrm{MSE}\left(\frac{\partial Y}{\partial h}\frac{\partial h}{\partial x_1},\hat{y}\right)$, where $\hat{y}=\partial\hat{Y}/\partial x_1$ is the derivative data.

Neural networks have the property of being uniform approximations of continuous functions over compact domains \cite{Cybenko1989}. However, underlying symmetries of the domain are not guaranteed to be reproduced exactly by the DNN. For instance, consider a function, $f(x,y)$, defined on variables $x$ and $y$, such that they are symmetric under the inversion operation. Given the symmetry of the domain of $(x,y)$, the function is also symmetric under the same operation, i.e. $f(x,y)=f(\pm x,\pm y)$. Any artificial neural network that approximates $f$ must reproduce this symmetry exactly. This can be enforced on the DNN by first transforming the inputs to a set of symmetric functions \cite{Ling2015,Ling2016}. For example, within a neural network to approximate $f$, rather than use the values of $(x,y)$ as input, the symmetric functions $(x^{2},y^{2})$ can be used to parameterize the DNN. These functions map all symmetrically equivalent points in the $(x,y)$ space on to the same value, while also differentiating symmetrically distinguishable points. As another example consider functions $f(x,y)$ that are invariant under an inversion about the $y=0$ line. A DNN that approximates any function in this class can be guaranteed to obey the required symmetry by using $(x,y^{2})$ as the input functions. \par
In general, a set of invariant inputs to the neural network may be defined by generating symmetry invariant polynomial functions with algorithms described by Thomas and Van der Ven \cite{thomas2017}. We will denote these functions as $h(\cdot)$. Since the given derivative data is likely differentiated with respect to, for example, $y$ and not $y^{2}$, we must incorporate the symmetrized inputs, $h(\cdot)$, into the loss function, e.g. $\mathrm{MSE}\left(\frac{\partial Y}{\partial h}\frac{\partial h}{\partial x_1},\hat{y}\right)$, where $\hat{y}=\partial\hat{Y}/\partial x_1$ is the derivative data.

\section{Bridging atomic to continuum scales via an IDNN representation of the free energy density}
\label{sec:B2}
\begin{figure}[tb]
  \centering
\begin{minipage}[t]{0.5\textwidth}
        \centering
    \includegraphics[scale=0.4]{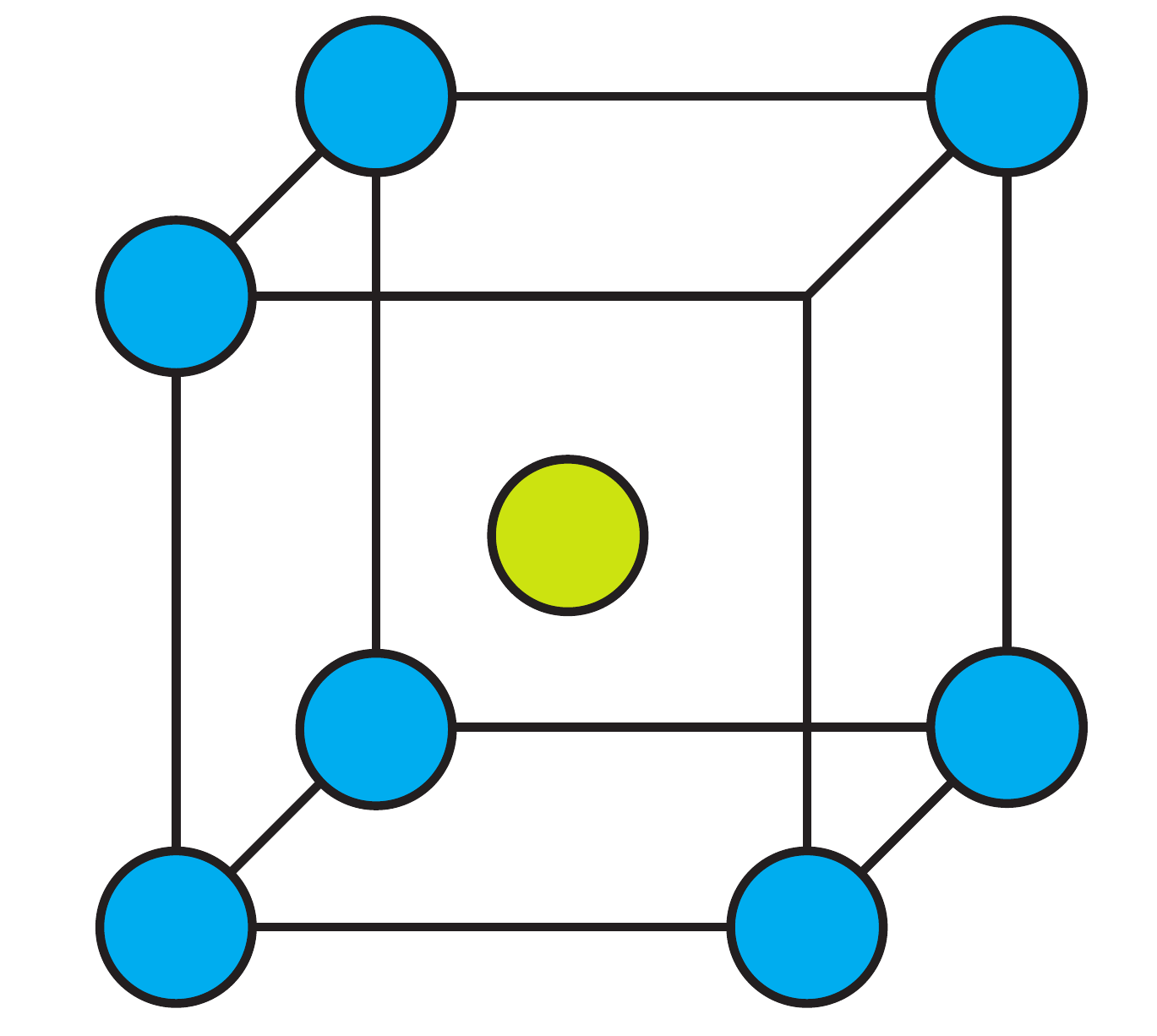}
	\subcaption{}
\end{minipage}%
\begin{minipage}[t]{0.5\textwidth}
        \centering
    	\includegraphics[scale=0.4]{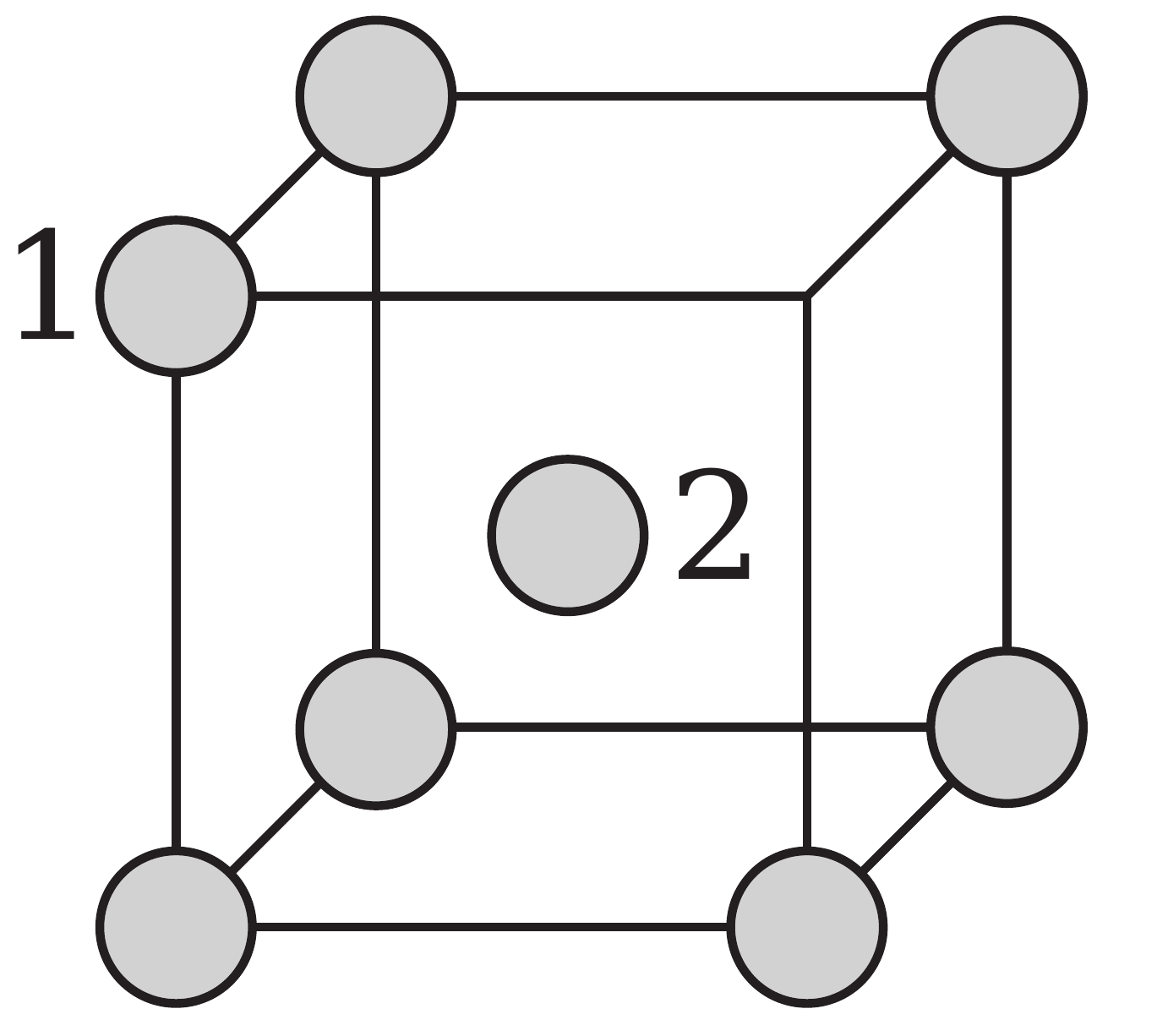}
	\subcaption{}
\end{minipage}
  \caption{(a) Schematic crystal structure of the B2 ordering of A (blue) and B (green) atoms on the bcc crystal structure (b) Conventional cell of bcc showing two sublattices labeled 1 and 2 that are used to calculate sublattice compositions.}
  \label{fig:bcc_conventional}
\end{figure}

The phase field model is an outgrowth of the Allen-Cahn theory that was originally developed to describe the motion of anti-phase boundaries in ordered compounds. 
Many binary alloys form a disordered solid solution at elevated temperature that transforms to an ordered compound at low temperature. 
A well-known example is the order-disorder transition in brass, where an equiatomic alloy of Cu and Zn forms a disordered mixture over the sites of a body centered cubic (bcc) lattice at high temperature, but then orders into a CsCl-type structure at low temperatures. 
The low temperature CuZn phase can be viewed as an ordered arrangement of Cu and Zn over the sites of the bcc lattice, as illustrated in Figure \ref{fig:bcc_conventional}.
This ordering on bcc is referred to as B2. 
Because of the translation symmetry of the bcc lattice, there are two ways to form distinct B2 orderings on the same parent bcc lattice.  
The Zn could, for example, occupy the corners of the conventional cubic bcc unit cell, while the Cu occupies the body centered sites. 
Alternatively, Cu could occupy the corners, while Zn occupies the body centered sites. 
The two orderings are identical, differing only by a rigid translation in space. 
An alloy that orders upon cooling may adopt the first variant in some regions of the solid and the second variant in other regions. 
When the two variants impinge, they are separated by an anti-phase boundary. 

Order parameters can be used to distinguish regions where the constituents of the alloy are ordered from regions where they are disordered. 
The order parameters should also be able to distinguish between the different translational variants of the ordered phase. 
Order parameters that satisfy these conditions for the B2 ordering on bcc are well known \cite{Allen1979,Natarajan2017}. 
They are defined as linear combinations of sublattice concentrations $x_1$ and $x_2$ that track the average concentrations over the two sublattice sites $1$ and $2$ of the cubic unit cell shown in Figure \ref{fig:bcc_conventional}. 
In a binary A-B alloy (e.g. A=Cu and B=Zn), each sublattice concentration $x_i$ ($i=1,2$) is defined as the fraction of crystal sites belonging to sublattice site $i$ that are occupied by B atoms. 
Convenient order parameters to measure the degree of B2 long-range ordering can be defined as \cite{Allen1979,Natarajan2017}
\begin{equation}
  \label{eq:homogeneous_comp}
  c = \frac{x_{1}+x_{2}}{\sqrt{2}}
\end{equation}
\begin{equation}
  \label{eq:op_B2}
  \eta = \frac{x_{1}-x_{2}}{\sqrt{2}}
\end{equation}
The first order parameter, $c$, is a measure of the homogeneous composition of the alloy (i.e. proportional to the fraction of all crystal sites occupied by B atoms).
The second order parameter, $\eta$, measures the degree of long-range B2-like ordering, and is equal to zero in the completely disordered state as the distinction between the two sublattices disappears in the absence of long-range order and $x_1$=$x_2$. The order parameter $\eta$ is also able to distinguish between the two translational variants of B2. This becomes clear when considering an alloy containing an equal number of A and B components (i.e. when $c=1/\sqrt{2}$). In one translational variant of B2, the B atoms occupy sublattice $1$ while the A atoms occupy sublattice $2$. For this variant $x_1=1$ and $x_2=0$ such that $\eta$ becomes equal to $1/\sqrt{2}$. In the second translational variant of B2, the site occupancies are reversed and $\eta$ becomes equal to $-1/\sqrt{2}$.\par

\begin{figure}[tb]
        \centering
    \includegraphics[scale=0.4]{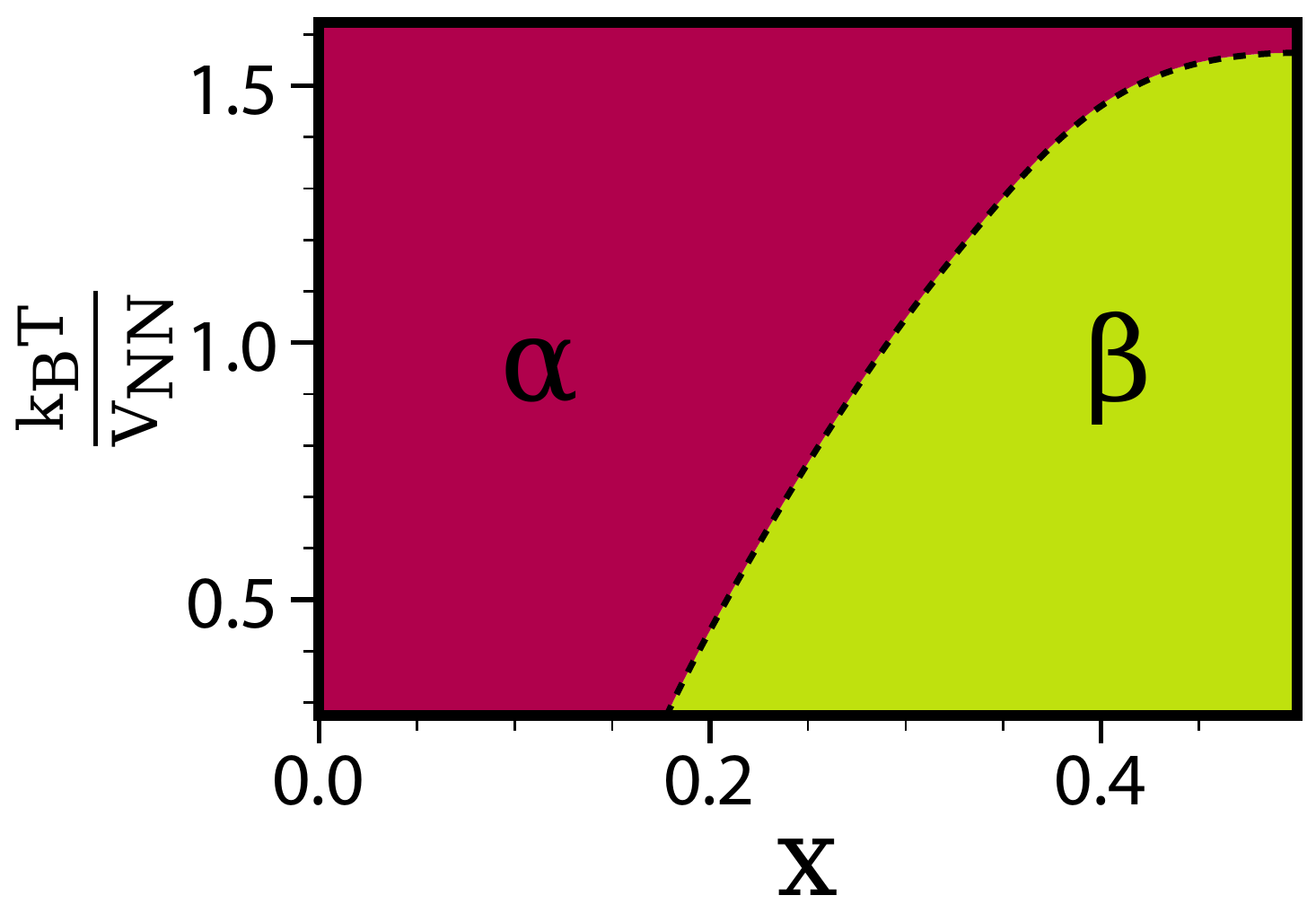}
\caption{Temperature-composition phase diagram calculated for the attractive nearest neighbor cluster expansion. Second order phase transitions are estimated from the divergence of the heat capacity. The temperature is normalized based on the nearest neighbor pair interaction (V$_{\textrm{NN}}$)}
\label{fig:clex_pd}
\end{figure}

The free energy, $g$, of a binary A-B alloy with a low-temperature preference for B2 ordering will be a function not only of temperature, $T$, and composition, $x$ (= $c/\sqrt{2}$), but also the long-range order parameter $\eta$. The dependence of $g$ on $T$, $x$ and $\eta$ can be calculated with statistical mechanics approaches. A binary alloy can be modeled as a lattice model that tracks the configurational degrees of freedom associated with all possible ways of arranging A and B atoms over $M$ sites of a crystal. This is done by assigning an occupation variable $\sigma_{i}$ to each crystal site $i=1,...,M$, which is equal to +1 when site $i$ is occupied by a B atoms and -1 when the site is occupied by an A atom. Any arrangement of A and B atoms on the M-site crystal is then fully specified by the configuration vector $\vec{\sigma}=\left(\sigma_1,...,\sigma_i,...,\sigma_M\right)$.
The energy of the crystal for any configurations $\vec{\sigma}$ can be described with a cluster expansion Hamiltonian \cite{Sanchez1984,deFontaine1994}
\begin{equation}
    E\left(\vec{\sigma}\right) = V_o+\sum_i V_i \sigma_i +\sum_{i,j}V_{i,j}\sigma_i \sigma_j+\sum_{i,j,k}V_{i,j,k}\sigma_i \sigma_j\sigma_k+...
\end{equation}
where the expansion coefficients $V_o$, $V_i$, $V_{i,j}$, $V_{i,j,k}$,... are determined by the chemistry of the alloy and can be parameterized by training to a database of first-principles electronic structure calculations. Here, we use a simplified model cluster expansion Hamiltonian that contains only the constant $V_o$, a point term, $V_i$ and a term for the nearest neighbor cluster $V_{i,j}$ where ${i,j}$ represents nearest neighbor clusters. The nearest neighbor pair interaction was chosen such that the B2 ordering is a ground state \cite{ducastelle1991}. 
The finite temperature thermodynamics associated with the order-disorder phase transition can be calculated with Monte Carlo simulations applied to the cluster expansion Hamiltonian. 
Figure \ref{fig:clex_pd} shows the temperature versus composition phase diagram of the model alloy. The B2 stability domain (green) is separated from the disordered solid-solution domain (crimson) by a second order phase transition (solid line).\par


Conventional grand-canonical Monte-Carlo techniques restrict the calculated free energy to only the thermodynamically stable regions. However, since the unstable parts of the free energy are critical to describing the temporal and spatial evolution of the system within a phase field model, we employed biased Monte-Carlo techniques to sample the free energy throughout the composition, order-parameter domain \cite{Natarajan2017}. 
The biased ensemble is defined with the following partition function:
\begin{equation}
  Z(\phi_{c},\phi_{\eta},\kappa_{c},\kappa_{\eta},T) = \sum_{\vec{\sigma}} \exp\left( - \frac{E(\vec{\sigma}) + M \phi_{c} (c(\vec{\sigma})-\kappa_{c})^{2} +  M \phi_{\eta} (\eta(\vec{\sigma})-\kappa_{\eta})^{2}}{k_{B} T} \right)
\end{equation}
where $E(\vec{\sigma})$, $c(\vec{\sigma})$ and $\eta(\vec{\sigma})$ are the energy, homogeneous composition and long-range order parameter evaluated for configuration $\vec{\sigma}$. The quantities $\phi_c$, $\phi_\eta$, $\kappa_c$ and $\kappa_\eta$ define bias potentials, with curvatures given by $\phi_c$ and $\phi_\eta$, and centers at $\kappa_c$ and $\kappa_\eta$, respectively. The biased ensemble is sampled with Metropolis Monte-Carlo, and statistical averages of the homogeneous composition $\langle c\rangle$, and order parameter $\langle\eta\rangle$ are measured for different values of $\phi$, $\kappa$, and $T$. The statistical averages can be related to the derivatives of the free energy per atom (denoted $g = G/M$, where $G$ is the total free energy) as \cite{Natarajan2017}:
\begin{equation}
  \label{chem_pot_derivative}
  \mu_{c} = \frac{\partial g}{\partial c}\Bigr|_{(\langle c \rangle,\langle \eta \rangle)} = -2 \phi_{c} (\langle c \rangle -\kappa_{c})
\end{equation}
\begin{equation}
  \label{op_derivative}
  \mu_{\eta}= \frac{\partial g}{\partial \eta}\Bigr|_{(\langle c \rangle,\langle \eta \rangle)} = -2 \phi_{\eta} (\langle \eta \rangle -\kappa_{\eta})
\end{equation}
where $\mu_{c}$ and $\mu_{\eta}$ are the chemical potentials with respect to the composition and order parameter respectively. The cluster expansions, and statistical mechanics calculations were performed with the \texttt{CASM} code \cite{Casm,Vanderven2018,thomas2013,Puchala2013}. The measured ensemble averages were then used to calculate free energy derivatives.

\subsection{Training free energy DNN}
\label{sec:training}

\begin{figure}[tb]
    \centering
    \includegraphics[width=0.6\textwidth]{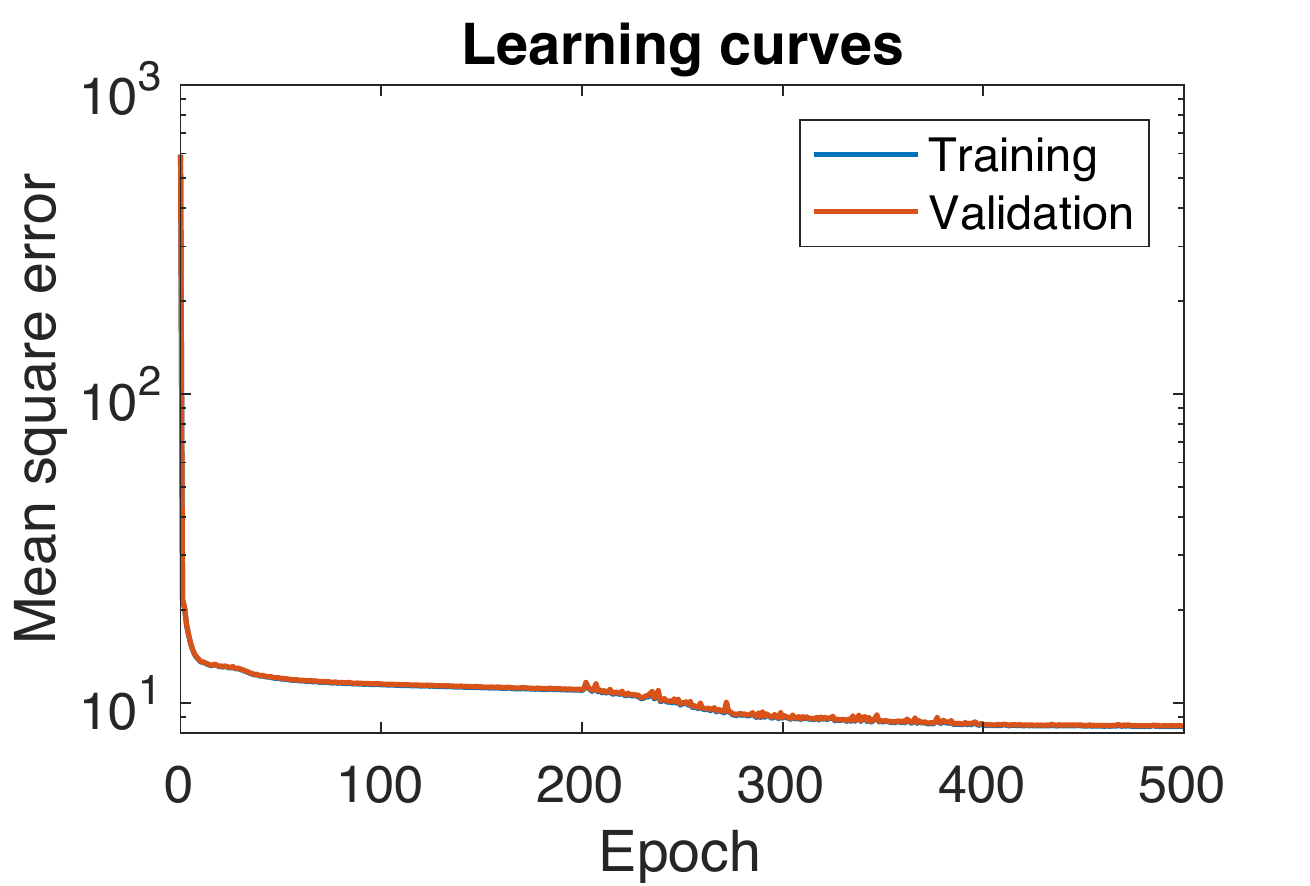}
    \caption{Learning curve for the IDNN showing the decrease in mean square error for testing and cross-validation datasets over training epochs. The cross-validation error is nearly indistinguishable from the training error.}
    \label{fig:learningCurve}
\end{figure}

An IDNN representing the chemical potential data, $\mu_c$ and $\mu_\eta$, was trained to the derivative data, i.e. the chemical potential values, as described in Section \ref{sec:DNN}, such that the IDNN could be analytically integrated to recover the free energy. The IDNN was implemented as a custom Estimator using the \texttt{TensorFlow} library \cite{Martin2015}, and was defined by two hidden layers with 10 units per layer. The IDNN was trained for 500 epochs using the \texttt{AdagradOptimizer}, with learning rates of 0.1 and 0.5 applied at different stages of training. A batch size of 10 was used, with 105,061 points in the training set and 35,021 points used for cross-validation. The resulting learning curve is plotted in Figure \ref{fig:learningCurve}, showing the decrease of both the training and cross-validation mean square errors as training progressed. Symmetry with respect to the order parameter about $\eta=0$ was enforced.

\begin{figure}[tb]
    \centering
    \includegraphics[width=0.6\textwidth]{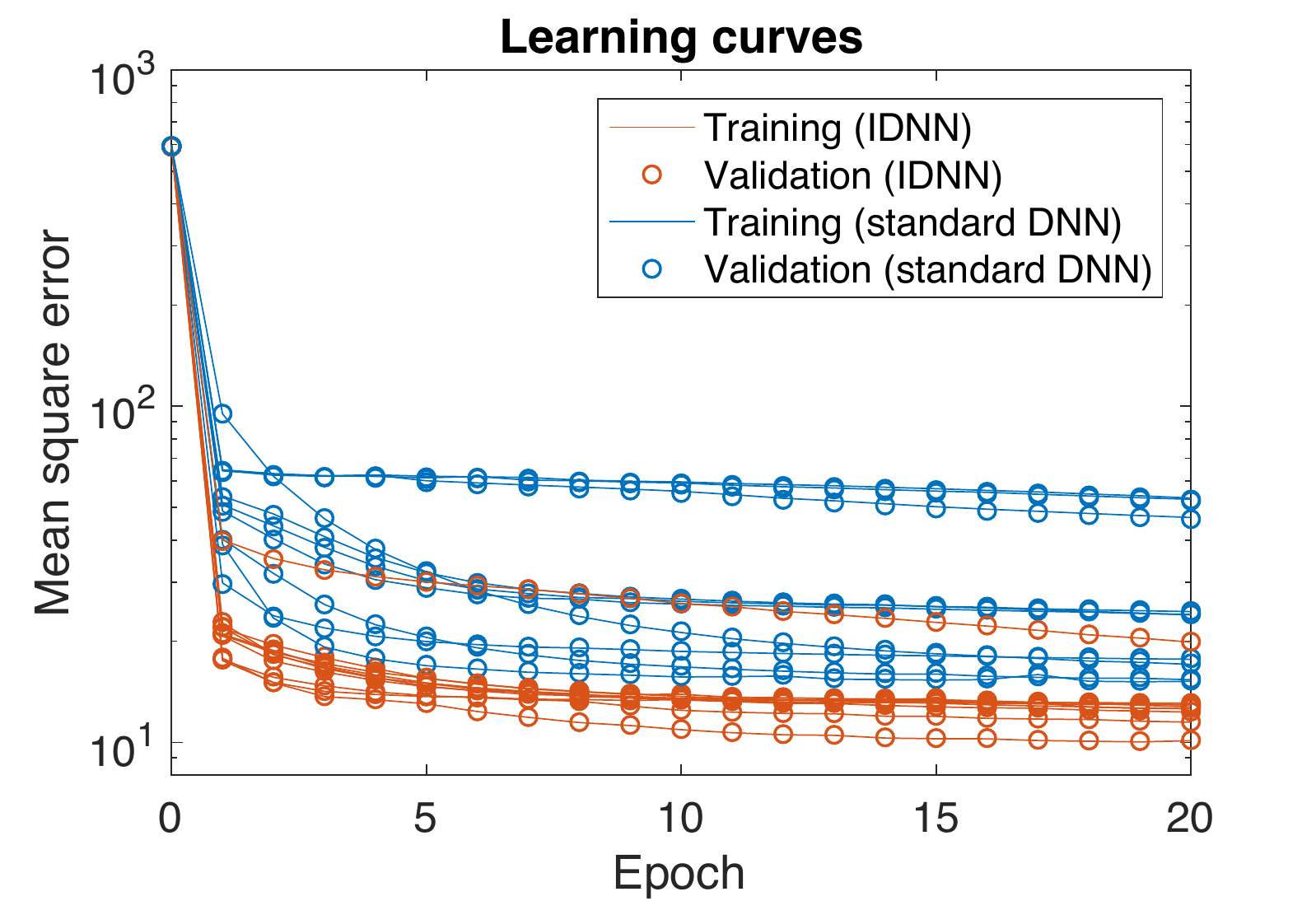}
    \caption{The training of ten standard DNNs is compared with the training of ten IDNNs. Each neural network was trained for 20 epochs, with different initial values for the weights and biases. The cross-validation error is nearly indistinguishable from the training error. The added complexity of the IDNN does not inhibit training.}
    \label{fig:learningCurveComp}
\end{figure}

Since the IDNN has a more complex form than the standard DNN, it is reasonable to expect that it may require more training to achieve comparable errors. To demonstrate any differences in training, ten IDNNs and ten standard DNNs were trained to the same chemical potential data, with the same symmetry conditions imposed. All twenty neural networks consisted of two hidden layers of ten neurons, each with different initial values for the weights and biases. They were trained for 20 epochs with a learning rate of 0.1. The resulting learning curves appear in Figure \ref{fig:learningCurveComp}. After 20 epochs, the average MSE for the standard DNNs was higher than the average MSE for the IDNNs. Thus, the added complexity of the IDNN does not inhibit the training.

\begin{figure}[tb]
        \centering
\begin{minipage}[t]{0.5\textwidth}
        \centering
	\includegraphics[width=0.95\textwidth]{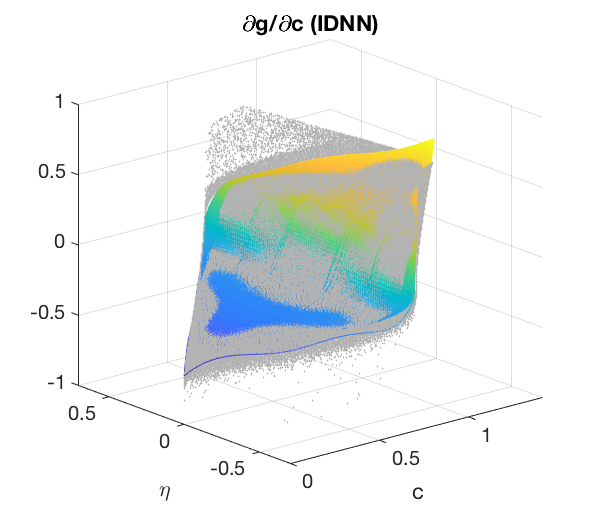}
% 	\caption{With $m=0.5$}
\end{minipage}%
\begin{minipage}[t]{0.5\textwidth}
        \centering
	\includegraphics[width=0.95\textwidth]{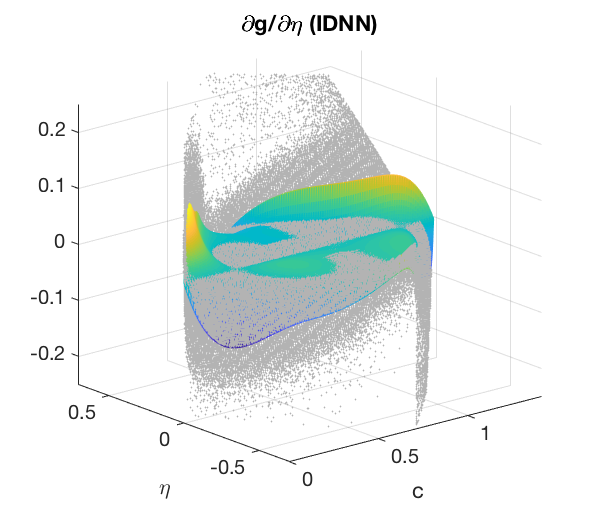}
% 	\caption{With $m=1.5$}
\end{minipage}
\begin{minipage}[t]{0.5\textwidth}
        \centering
	\includegraphics[width=0.95\textwidth]{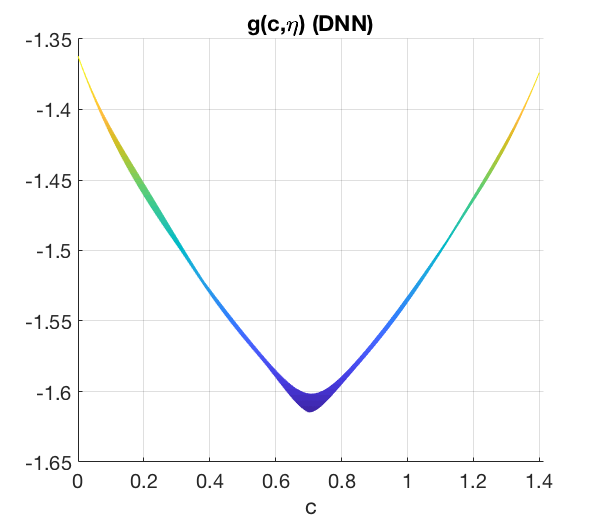}
% 	\caption{With $m=0.5$}
\end{minipage}%
\begin{minipage}[t]{0.5\textwidth}
        \centering
	\includegraphics[width=0.95\textwidth]{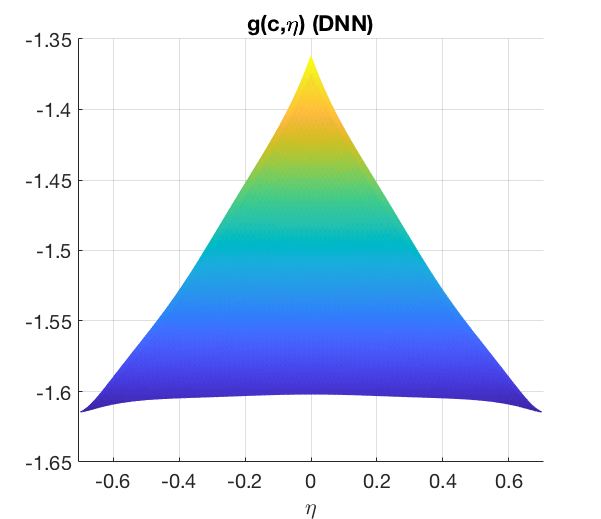}
% 	\caption{With $m=1.5$}
\end{minipage}
\caption{Top row: plots of the chemical potential IDNN representations (surfaces) with the chemical potential data points (grey). Bottom row: the analytically integrated free energy density DNN.}
\label{fig:dnn_fe}
\end{figure}

Figure \ref{fig:dnn_fe} shows the original chemical potential data compared with the associated IDNN that was trained to the corresponding chemical potential data. It also shows two views of the free energy surface as represented by the analytically integrated DNN. Perhaps the most significant feature of the free energy surface is the two energy wells, located at about $c=1/\sqrt{2}$, $\eta = \pm 1/\sqrt{2}$. Given that the wells exist at the same composition, the material will not separate into multiple phases, but instead form anti-phase domains. This reflects the expected physics of the system, described at the beginning of Section \ref{sec:B2}.

\subsection{Phase field computation}
\label{sec:phase field}
To demonstrate the use of the DNN representation of the free energy in computations, the analytically integrated free energy DNN was used in phase field computations. The phase field model was based on the coupled Cahn-Hilliard and Allen-Cahn equations \cite{CahnHilliard1958,Allen1979,Cahn1996,NovickCohen2000,Barrett2001}, solved using isogeometric analysis (IGA) \cite{CottrellHughesBazilevs2009}. The simulation was performed using the \texttt{mechanoChemIGA} code\footnote{Code available at github.com/mechanoChem/mechanoChemIGA}, which is based on the \texttt{PetIGA} \cite{Dalcin2016} and \texttt{PETSc} \cite{petsc-web-page,petsc-user-ref,petsc-efficient} libraries, and run on the XSEDE Comet HPC cluster \cite{Towns2014}.

\subsubsection{Formulation}
Given the homogeneous free energy density $g(c,\eta)$ as a function of concentration, $c$, and order parameter, $\eta$, we define the total free energy as the following:
\begin{align}
    \Pi[c,\eta] &= \int\limits_\Omega \left[g(c,\eta) + \frac{1}{2}\chi|\nabla c|^2 + \frac{1}{2}\chi|\nabla\eta|^2\right]\,\mathrm{d}V
\end{align}
% where the homogeneous free energy can be additively decomposed into the ideal solution free energy and the excess free energy, as in $g(c,\eta) = \hat{g}(c,\eta) + \Delta g(c,\eta)$, respectively.
The corresponding chemical potentials are given by the variational derivatives of the total free energy, namely $\mu_c := \delta\Pi/\delta c$ and $\mu_\eta := \delta\Pi/\delta\eta$. Using standard variational methods results in the following equations for the chemical potentials:
% \begin{align}
%     \mu_c &= \frac{1}{2}\frac{k_BT}{\sqrt{2}}\log\left(\frac{c^2-\eta^2}{(\sqrt{2}-c)^2-\eta^2}\right) + \frac{\partial \Delta g}{\partial c} - \chi_1 \nabla^2 c\\
%     \mu_\eta &= \frac{1}{2}\frac{k_BT}{\sqrt{2}}\log\left(\frac{(c+\eta)(\sqrt{2}-c+\eta)}{(c-\eta)(\sqrt{2}-c-\eta)}\right) + \frac{\partial \Delta g}{\partial \eta} - \chi_2 \nabla^2 \eta
% \end{align}
\begin{align}
    \mu_c &= \frac{\partial g}{\partial c} - \chi \nabla^2 c\\
    \mu_\eta &= \frac{\partial g}{\partial \eta} - \chi \nabla^2 \eta
\end{align}

The phase field model consists of the Cahn-Hilliard and Allen-Cahn equations, given by the following, respectively:
\begin{align}
    \frac{\partial c}{\partial t} &= 4\chi\nabla\cdot M(c,\eta)\nabla\mu_c\\
    \frac{\partial \eta}{\partial t} &= -\frac{1}{4}M(c,\eta)\mu_\eta
\end{align}
The Cahn-Hilliard equation is in conservation form, with the flux defined as $\bsym{J} := -M\nabla\mu_c$. It models the overall composition of the system through $c$, while conserving mass. The Allen-Cahn equation models the time evolution of the long-range ordering of the system through the non-conserved order parameter $\eta$. The two equations are coupled through the chemical potentials being derived from the same free energy. Periodic boundary conditions were applied. A degenerate mobility of the following form was applied:
\begin{equation}
    M(c,\eta) = 16\tilde{M}c^2(\sqrt{2}-c)^2(1/2 - \eta^2)^2
\end{equation}
% We take the following conditions on the boundary $\partial\Omega$:
% \begin{align}
%     M\nabla\mu_c\cdot\bsym{n} &= J_n\\
%     \nabla \eta\cdot\bsym{n} &= 0\label{eqn:grad eta}\\
%     \nabla c\cdot\bsym{n} &= 0\label{eqn:grad c}
% \end{align}
% where the conditions in Equations (\ref{eqn:grad eta}) and (\ref{eqn:grad c}) arise from enforcing equilibrium at the boundary.

The weak form of the equations, as solved by the IGA formulation, takes the following form:\begin{align}
\begin{split}
    0 &= \int_\Omega \left[w\frac{\partial c}{\partial t} + 4\chi \left(M\nabla w\cdot\nabla g_{,c}+\chi(M\nabla^2w + \nabla M\cdot\nabla w) \nabla^2 c\right)\right]\mathrm{d}V
\end{split}\\
\begin{split}
    0 &= \int_\Omega \left[w\frac{\partial \eta}{\partial t} + \frac{1}{4}\left(Mwg_{,\eta} + \chi(M\nabla w + w\nabla M)\cdot\nabla\eta\right)\right]\mathrm{d}V
\end{split}
\end{align}
Due to the stiffness of the equations when using the realistic free energy density, a two-stage, fourth-order Runge-Kutta time stepping scheme was used \cite{Suli2003}.
% \begin{align}
% \begin{split}
%     0 &= \int_\Omega \left[w\frac{\partial c}{\partial t} + M\left(\nabla w\cdot\nabla g_{,c}+\chi_1\nabla^2w\nabla^2c\right)\right]\,\mathrm{d}V\\
%     &\phantom{=} - \int_{\partial\Omega} [wJ_n + M\chi_1(\nabla^2c(\nabla w\cdot\bsym{n})+\nabla^2w(\nabla c\cdot\bsym{n}))]\,\mathrm{d}S\\
%     &\phantom{=}+ \int_{\partial\Omega}\tau(\nabla w\cdot\bsym{n})(\nabla c\cdot\bsym{n})\,\mathrm{d}S\label{eq:CHNitsche}
% \end{split}\\
% \begin{split}
%     0 &= \int_\Omega \left[w\frac{\partial \eta}{\partial t} + L(wg_{,\eta} + \chi_2\nabla w\cdot\nabla\eta))\right]\,\mathrm{d}V\\
%     %&\phantom{=}- \int_{\partial\Omega}L\chi_2[w\nabla\eta\cdot\bsym{n}+\eta\nabla w\cdot\bsym{n} - \tau(\nabla\eta\cdot\bsym{n})(\nabla w\cdot\bsym{n})]\,\mathrm{d}S
% \end{split}
% \end{align}
% where the higher order Dirichlet boundary condition in Equation (\ref{eqn:grad c}) is applied using Nitsche's method \cite{nitsche1971,arnold2002,Rudrarajuetal2016} in the last term of Equation (\ref{eq:CHNitsche}).

\subsubsection{Phase field results}
\label{sec:pfresults}
We considered a two-dimensional domain, discretized by a $200\times200$ element mesh. The initial and boundary value problem was initialized with a uniform composition field of $c = 1/\sqrt{2}$ and an order parameter field randomly perturbed about $\eta = 0$ by a value up to 0.01, representing a material that had just been quenched from a higher temperature. Periodic boundary conditions were applied. The initial time step was $\Delta t = 1$, with an adaptive time stepping scheme being applied to modify the time step based on the convergence of the nonlinear solver at each time step.

\begin{figure}[tb]
        \centering
\begin{minipage}[t]{0.5\textwidth}
        \centering
	\includegraphics[trim=2cm 4cm 2cm 1cm, clip,width=0.9\textwidth]{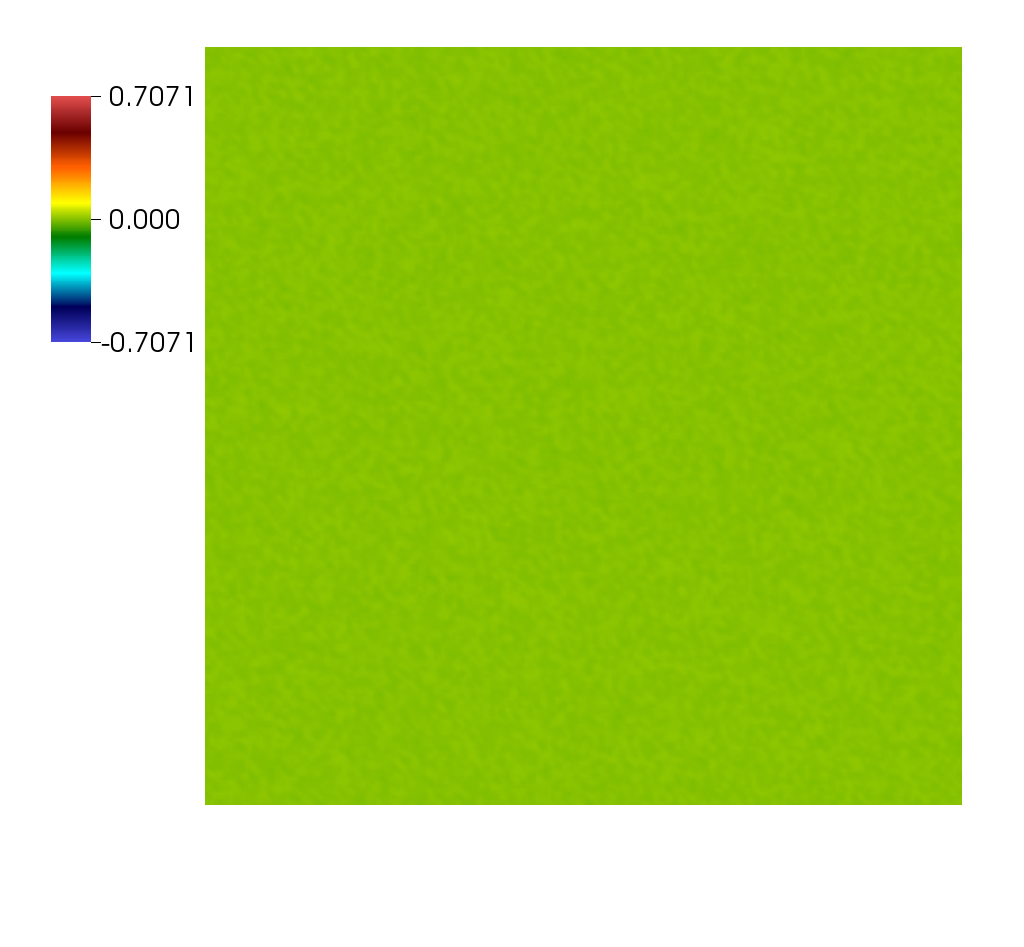}
	\subcaption{Time step 0 ($t=0$)}
\end{minipage}%
\begin{minipage}[t]{0.5\textwidth}
        \centering
	\includegraphics[trim=2cm 4cm 2cm 1cm, clip,width=0.9\textwidth]{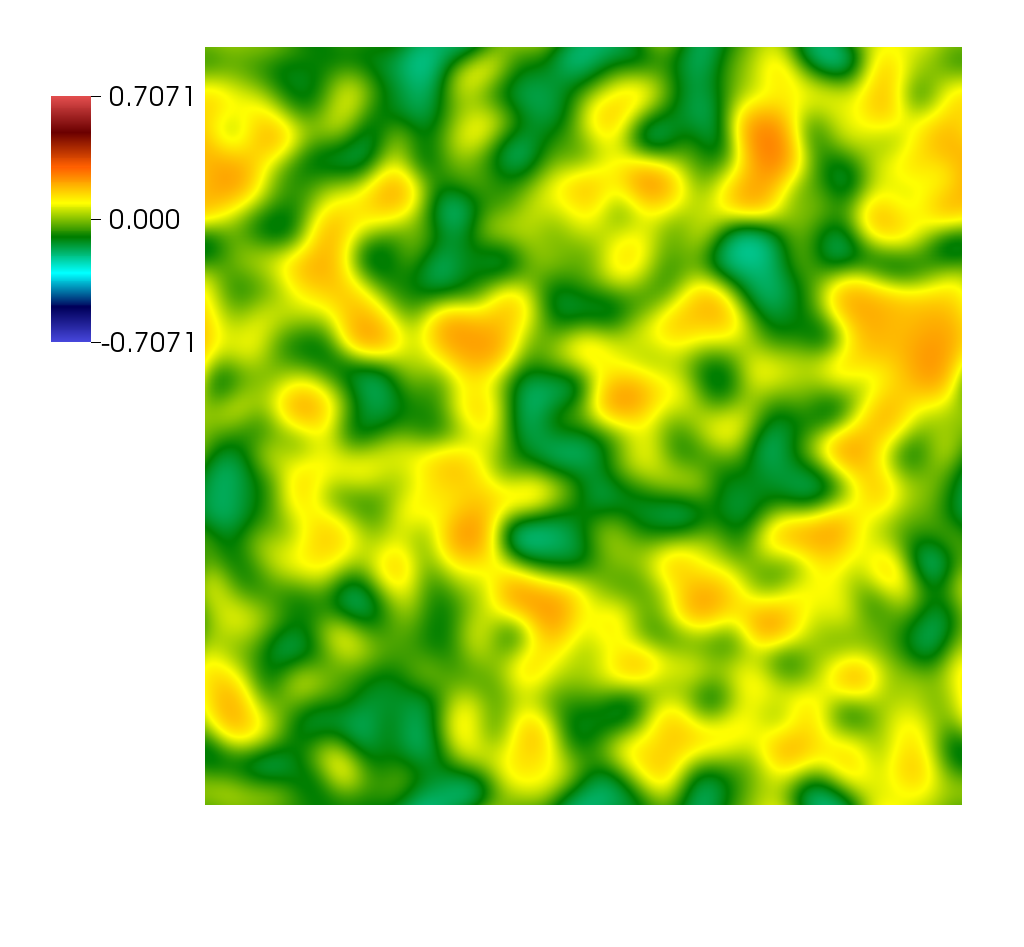}
	\subcaption{Time step 20 ($t=39.2$)}
\end{minipage}
\begin{minipage}[t]{0.5\textwidth}
        \centering
	\includegraphics[trim=2cm 4cm 2cm 1cm, clip,width=0.9\textwidth]{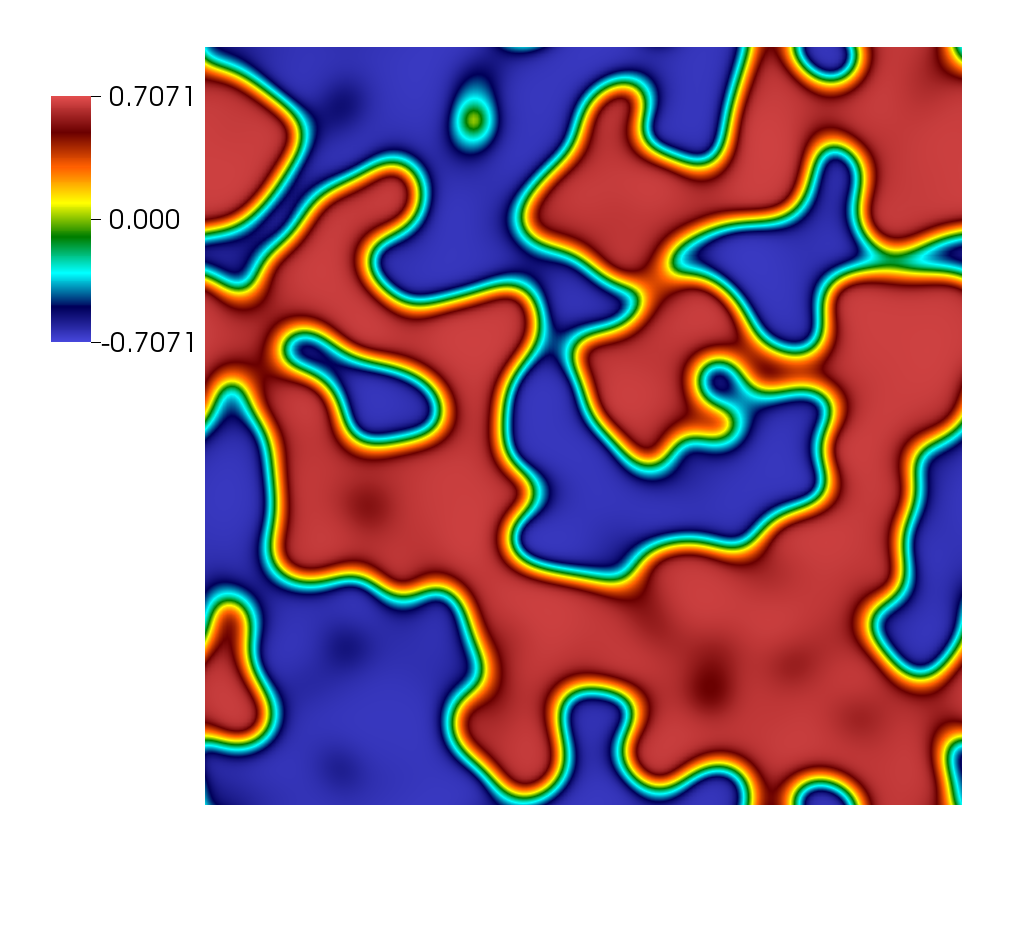}
	\subcaption{Time step 30 ($t=91.5$)}
\end{minipage}%
\begin{minipage}[t]{0.5\textwidth}
        \centering
	\includegraphics[trim=2cm 4cm 2cm 1cm, clip,width=0.9\textwidth]{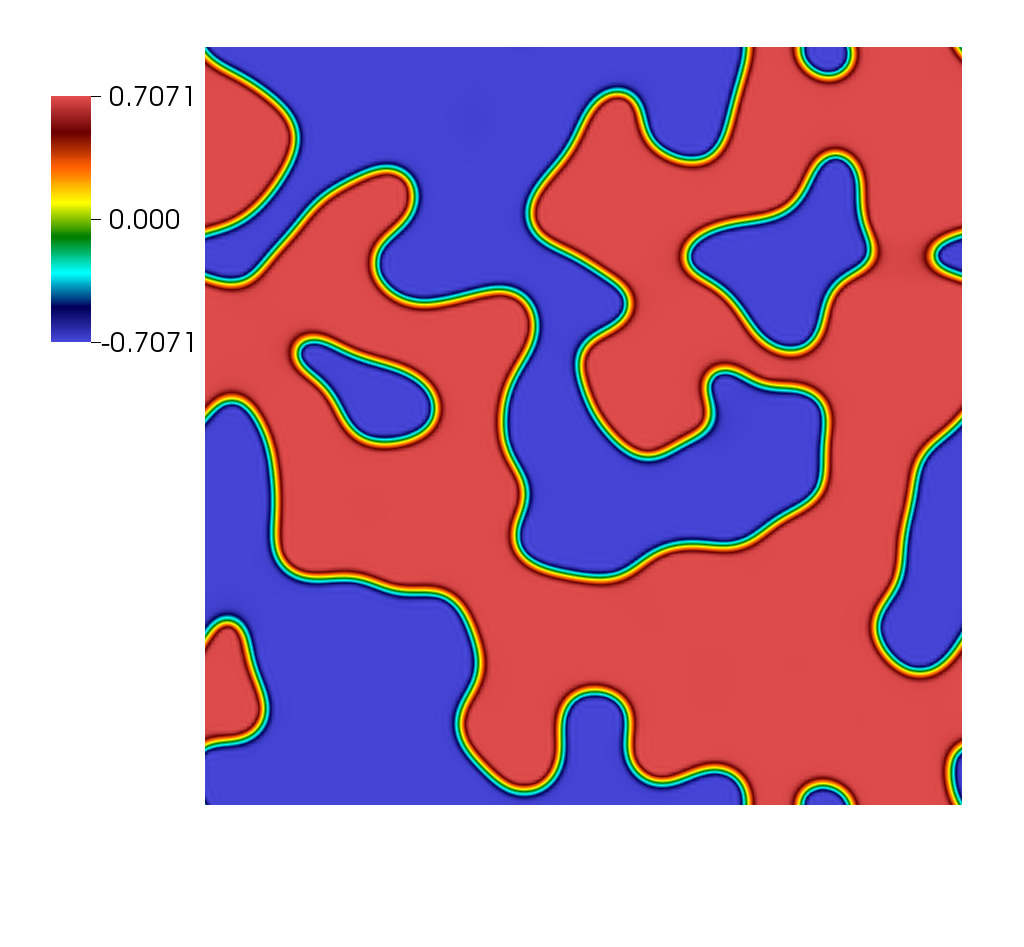}
	\subcaption{Time step 40 ($t=171.5$)}
\end{minipage}
\caption{The order parameter field from the phase field simulation is plotted, showing the formation of antiphase boundaries in a B2 alloy, with chemical potentials represented as IDNNs, integrated to yield the free energy.}
\label{fig:dnn_pf1}
\end{figure}

As shown in Figure \ref{fig:dnn_pf1}, the initially disordered domain gradually forms regions of the two ordered states. The antiphase boundary has completely formed within 40 time steps. The antiphase domains take on order parameter values of $\pm 1/\sqrt{2}$, while the composition field remains nearly uniform.

\begin{figure}[tb]
        \centering
\begin{minipage}[t]{0.5\textwidth}
        \centering
	\includegraphics[trim=2cm 4cm 2cm 1cm, clip,width=0.9\textwidth]{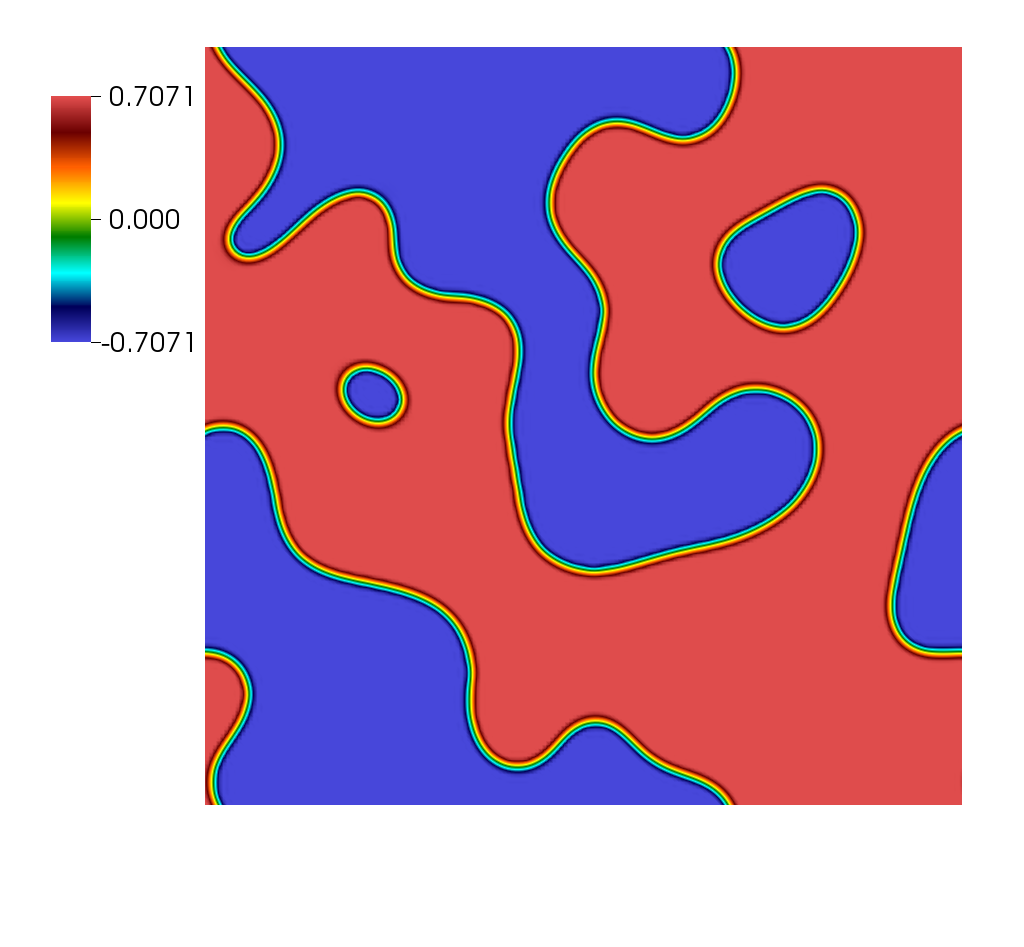}
	\subcaption{Time step 150 ($t=907$)}
\end{minipage}%
\begin{minipage}[t]{0.5\textwidth}
        \centering
	\includegraphics[trim=2cm 4cm 2cm 1cm, clip,width=0.9\textwidth]{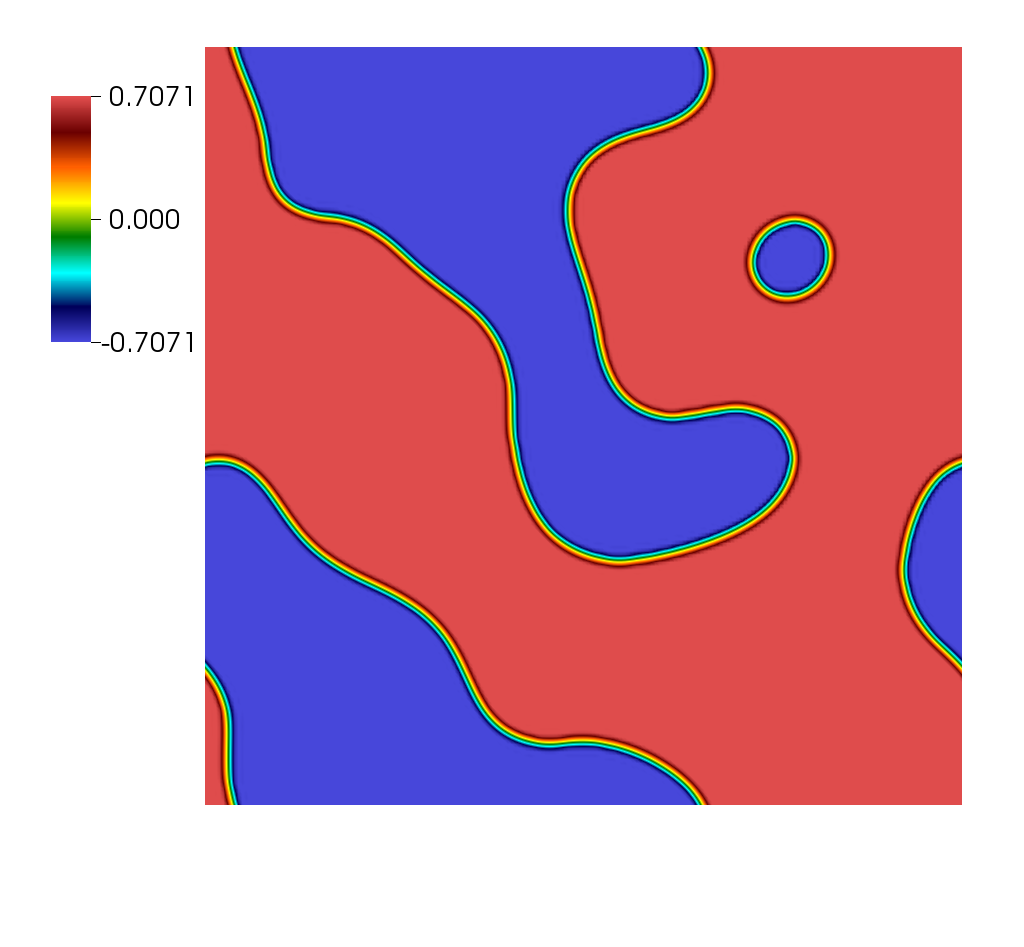}
	\subcaption{Time step 300 ($t=2154$)}
\end{minipage}
\begin{minipage}[t]{0.5\textwidth}
        \centering
	\includegraphics[trim=2cm 4cm 2cm 1cm, clip,width=0.9\textwidth]{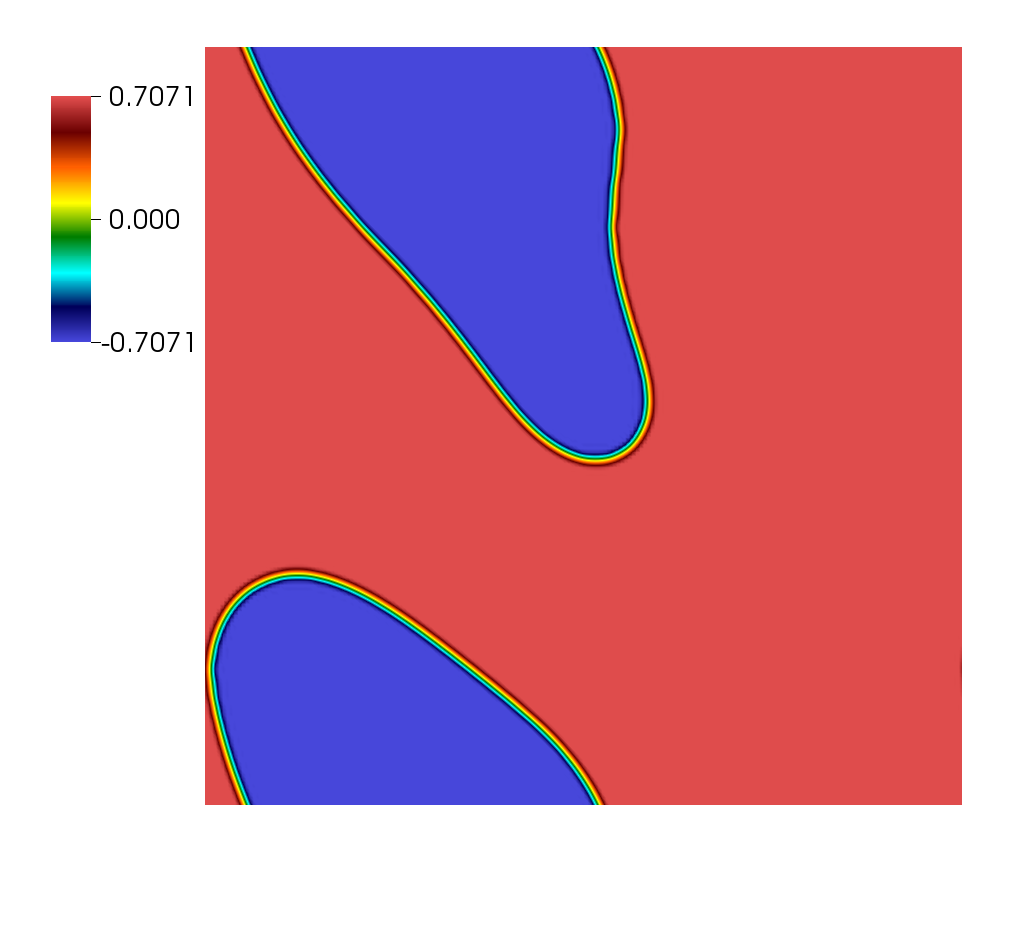}
	\subcaption{Time step 800 ($t=9725$)}
\end{minipage}
\caption{Curvature fluctuations away from straight antiphase boundaries decrease, and the domains coarsen with time. The chemical potentials are represented as IDNNs, integrated to yield the free energy.}
\label{fig:dnn_pf2}
\end{figure}

As the simulation progresses, the microstructure coarsens with  curvature fluctuations from a straight antiphase boundary decreasing, as shown in Figure \ref{fig:dnn_pf2}.

\section{Comparison with B-spline surface fit}
\label{sec:splines}
For comparison with the IDNN's representation properties of complex surfaces, we consider two-dimensional B-splines, motivated by their wide use in mathematics, and diverse applications in engineering \cite{CottrellHughesBazilevs2009}.

\subsection{Formulation}
The equation for a B-spline surface $Y(\xi_1,\xi_2)$, with knots on a tensor grid can be written in the following form
\begin{equation}
    Y(\xi_1,\xi_2) = \sum_{i=1}^m\sum_{j=1}^n C_{ij}N_{i,p}(\xi_1)M_{j,p}(\xi_2)
    \label{eqn:splinesurf}
\end{equation}
where $\xi_1\in[\xi^1_1,\xi^{m+p+1}_1]$, $\xi_2\in[\xi^1_2,\xi^{n+p+1}_2]$, and $N_{i,p}$ is the B-spline basis function of order $p$. The basis functions are defined by the Cox-de Boor recursion formula \cite{cox1972,deboor1972}
\begin{align}
    N_{i,p}(\xi_1) &= \frac{\xi_1 - \xi^i_1}{\xi^{i+p}_1-\xi^i_1}N_{i,p-1}(\xi_1)+\frac{\xi^{i+p+1}_1 - \xi_1}{\xi^{i+p+1}_1-\xi^{i+1}_1}N_{i+1,p-1}(\xi_1)\label{eq:coxdeboor1}\\
    N_{i,0}(\xi_1) &= 
    \begin{cases}
        1 & \text{if } \xi^i_1 \leq \xi_1 < \xi^{i+1}_1\\
        0 & \text{otherwise}
    \end{cases}
\end{align}
using the knot vector $\Xi_1 = \{\xi^1_1,\xi^2_1,\ldots,\xi^{m+p+1}_1\}$. $M_{j,p}(\xi_2)$ is similarly defined using the knot vector $\Xi_2 = \{\xi^1_2,\xi^2_2,\ldots,\xi^{n+p+1}_2\}$.

To convert the matrix $C$ to a vector (to use the standard form for least squares fitting), we use the following index conversion: $I = ni+j$, $i=0,\ldots,m-1$, $j=0,\ldots,n-1$, so that $I = 0,\ldots,mn-1$. Then we can rewrite Eq. (\ref{eqn:splinesurf}) as the following:

\begin{equation}
    Y(\xi_1,\xi_2) = \sum_{I=0}^{mn-1}c_IP_{I,p}(\xi_1,\xi_2)
\end{equation}
where $c_I := C_{ij}$ and $P_{I,p}(\xi_1,\xi_2) := N_{i,p}(\xi_1)M_{j,p}(\xi_2)$. If evaluating multiple data points $\{(\hat{\xi}_{1_k},\hat{\xi}_{2_k})\}$, we can write the resulting vector of function evaluations using the following matrix-vector form, written in coordinate notation:
\begin{align}
    Y_k &= A_{kI}c_I
\end{align}
where
\begin{equation}
    A_{kI} = P_{I,p}(\hat{\xi}_{1_k},\hat{\xi}_{2_k})
\end{equation}

We now consider fitting to two sets of derivative data. For derivative datasets contained in the two vectors $\bsym{\hat{\mu}}_1$ and $\bsym{\hat{\mu}}_2$, the following matrices are defined:
\begin{align}
    B_{kI} &= \frac{\partial}{\partial \xi_1}P_{I,p}(\hat{\xi}_{1_k},\hat{\xi}_{2_k})\\
    C_{kI} &= \frac{\partial}{\partial \xi_2}P_{I,p}(\hat{\xi}_{1_k},\hat{\xi}_{2_k})\label{eq:splineparam}
\end{align}
Then, we have the following least squares formulation, with some regularization added for numerical stability:
\begin{align}
    \hat{\bsym{c}} &= \mathrm{arg}\min_{\bsym{c}} \left[\left(\bsym{B}\bsym{c} - \bsym{\hat{\mu}}_1\right)^T\left(\bsym{B}\bsym{c} - \bsym{\hat{\mu}}_1\right) + \left(\bsym{C}\bsym{c} - \bsym{\hat{\mu}}_2\right)^T\left(\bsym{C}\bsym{c} - \bsym{\hat{\mu}}_2\right) + \lambda\bsym{c}^T\bsym{c}\right]
\end{align}
where $\lambda$ is the regularization coefficient.
% Continuing with coordinate notation for clarity, we take the gradient with respect to $\bsym{c}$ and set each component equal to zero:
% \begin{align}
%     \implies 0 &= \frac{d}{dc_L}\left[\left(B_{kI}c_I - \hat{\mu}_{1_k}\right)\left(B_{kJ}c_J - \hat{\mu}_{1_k}\right) + \left(C_{kI}c_I - \hat{\mu}_{2_k}\right)\left(C_{kJ}c_J - \hat{\mu}_{2_k}\right) +\lambda c_kc_k\right]\\
%     &= 2\left(B_{kI}c_I - \hat{\mu}_{1_k}\right)B_{kL} + 2\left(C_{kI}c_I - \hat{\mu}_{2_k}\right)C_{kL} + 2\lambda c_L\\
%     &= 2\left[\left(B_{kL}B_{kI}+ C_{kL}C_{kI} + \lambda\delta_{LI}\right)c_I - B_{kL}\hat{\mu}_{1_k} - C_{kL}\hat{\mu}_{2_k}\right]
% \end{align}
% where $\delta_{LI}$ is the Kronecker delta.
Setting the gradient with respect to $\bsym{c}$ equal to the zero vector leads to the following least squares solution:
\begin{align}
    \hat{\bsym{c}} &= \left(\bsym{B}^T\bsym{B}+\bsym{C}^T\bsym{C} + \lambda\bsym{I}\right)^{-1}\left(\bsym{B}^T\bsym{\hat{\mu}}_1+\bsym{C}^T\bsym{\hat{\mu}}_2\right)
    \label{eq:optimspline}
\end{align}
where $\bsym{I}$ is the identity matrix. Equations (\ref{eq:coxdeboor1}-\ref{eq:splineparam}) and \eqref{eq:optimspline} are applied, with datasets in the $(c,\eta)$ space corresponding to the $(\xi_1,\xi_2)$ space for B-spline surfaces. Chemical potential data for $\mu_c$ and $\mu_\eta$ are contained in the vectors $\hat{\bsym{\mu}}_1$ and $\hat{\bsym{\mu}}_2$, respectively.

\subsection{Selection of knots}
While the method presented in the previous section can be used to optimize the values of the control points for given knot vectors, the locations of the knots also can be optimized to minimize the error. A variety of approaches are available, including nonlinear least squares, bisection, and genetic algorithms \cite{Dung2017}. Our approach was performed in two steps.

% \begin{figure}[tb]
%         \centering
% \begin{minipage}[t]{0.5\textwidth}
%         \centering
% 	\includegraphics[width=0.99\textwidth]{figures/uniform_knots.pdf}
% 	\subcaption{Uniform}
% \end{minipage}%
% \begin{minipage}[t]{0.5\textwidth}
%         \centering
% 	\includegraphics[width=0.99\textwidth]{figures/Chebyshev_knots.pdf}
% 	\subcaption{Chebyshev}
% \end{minipage}
% \caption{The mean square error (MSE) is compared for different numbers of knots distributed both uniformly and using Chebyshev nodes.}
% \label{fig:knot_opt}
% \end{figure}

\begin{figure}[tb]
        \centering
	\includegraphics[width=0.6\textwidth]{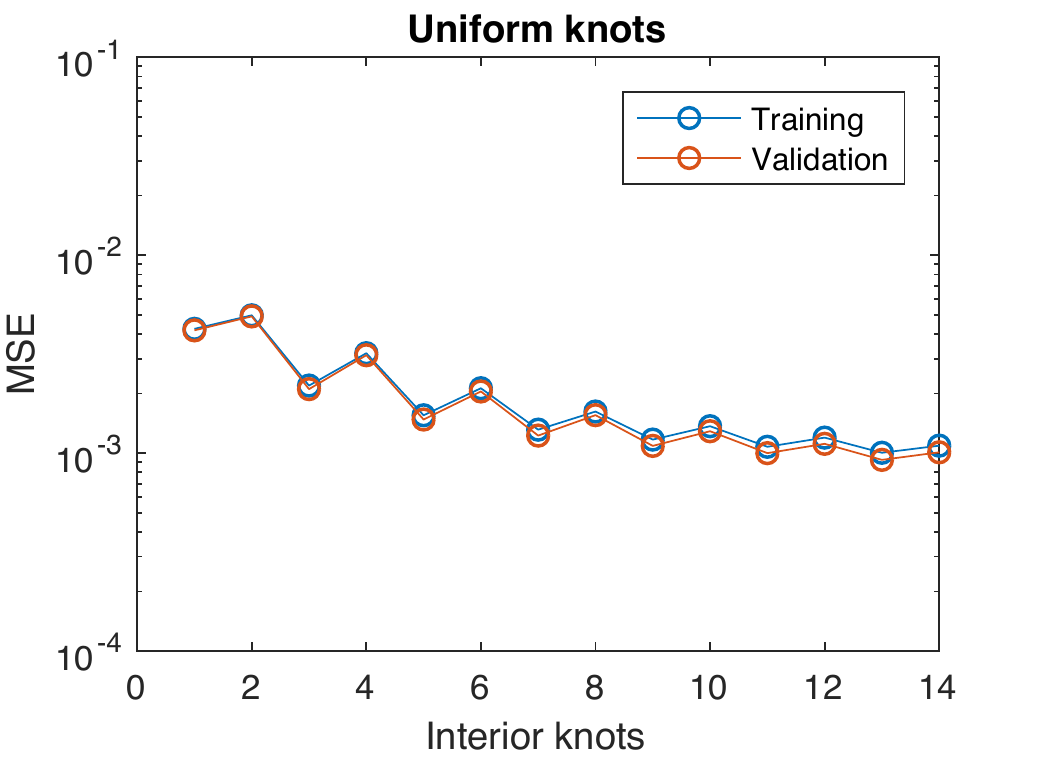}
\caption{The mean square error (MSE) is compared for different numbers of knots distributed uniformly.}
\label{fig:knot_opt}
\end{figure}

First, we divided the data into training data (75\%) and validation data (25\%). As before, symmetry of the function about $\eta=0$ was imposed. The optimal control points were found for a number of uniformly spaced knots, and the mean square error (MSE) using the validation data was reported, as shown in Figure \ref{fig:knot_opt}. Higher numbers of knots were not used due to the resulting oscillatory behavior of the fit, particularly in regions of missing data. Knots placed at Chebyshev nodes also resulted in inaccurate fluctuations where data was missing when more than one interior knot was used. In comparing the resulting cross-validation error, it was found that thirteen uniformly distributed interior knots gave the lowest error with $9.25\times10^{-4}$.

In the second step, we used Matlab's genetic algorithm optimization routine to attempt to improve the B-spline fit using thirteen interior knots per knot vector (26 total variables). Appropriate inequality constraints were applied to maintain monotonically increasing knot vectors. The algorithm terminated after 50 generations, with a MSE of $5.903\times10^{-3}$. Since this was not an improvement over the error with uniformly spaced knots, the B-spline fit with thirteen uniformly spaced interior knots per knot vector was taken as the best fit using B-splines.

\subsection{B-spline results, and comparison with DNN}
\begin{figure}[tb]
        \centering
\begin{minipage}[t]{0.5\textwidth}
        \centering
	\includegraphics[width=0.99\textwidth]{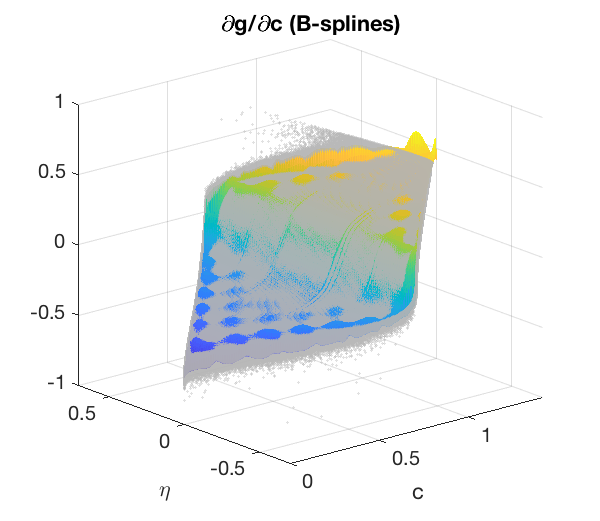}
% 	\subcaption{Chemical potential with respect to $c$}
\end{minipage}%
\begin{minipage}[t]{0.5\textwidth}
        \centering
	\includegraphics[width=0.99\textwidth]{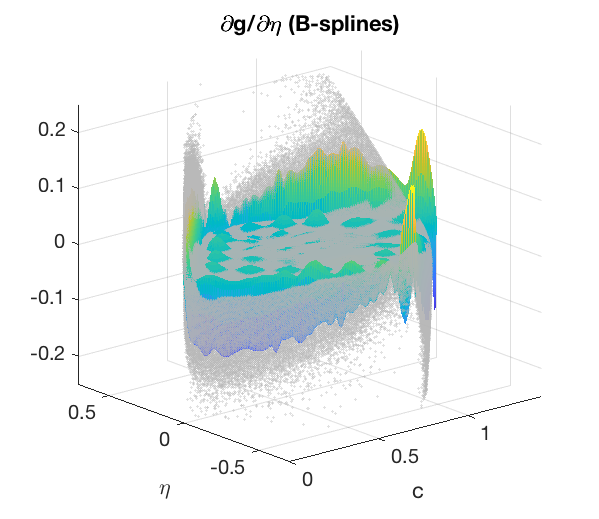}
% 	\subcaption{Chemical potential with respect to $\eta$}
\end{minipage}
\begin{minipage}[t]{0.5\textwidth}
        \centering
	\includegraphics[width=0.99\textwidth]{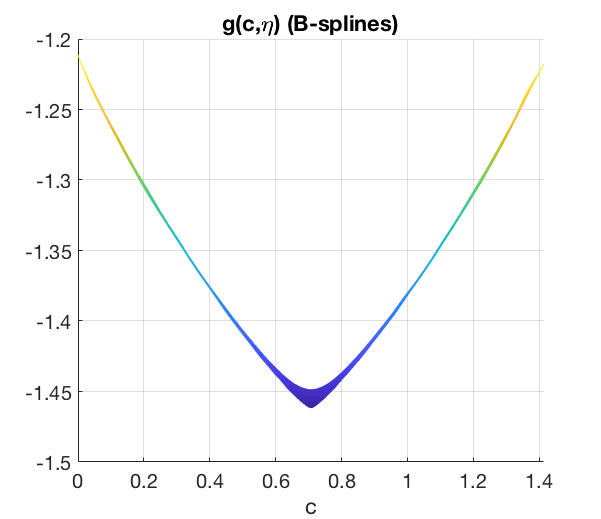}
% 	\subcaption{Free energy density, viewed along $\eta$-axis.}
\end{minipage}%
\begin{minipage}[t]{0.5\textwidth}
        \centering
	\includegraphics[width=0.99\textwidth]{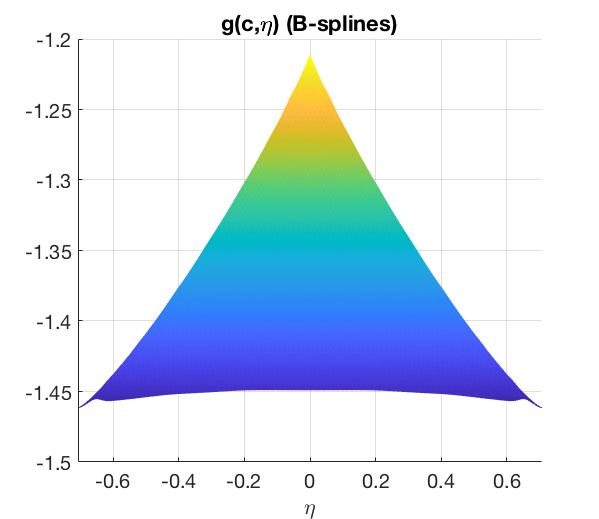}
% 	\subcaption{Free energy density, viewed along $c$-axis.}
\end{minipage}
\caption{Top row: Plots of the chemical potential B-spline surfaces with data in grey. Bottom row: Free energy density B-spline surface.}
\label{fig:spline_fe}
\end{figure}

The resulting B-spline representations of the chemical potentials and the free energy density are plotted in Figure \ref{fig:spline_fe}. Visually, these fits seem similar to those of the DNN in Figure \ref{fig:dnn_fe}, although there are notable differences.

\begin{figure}[tb]
        \centering
	\includegraphics[width=0.55\textwidth]{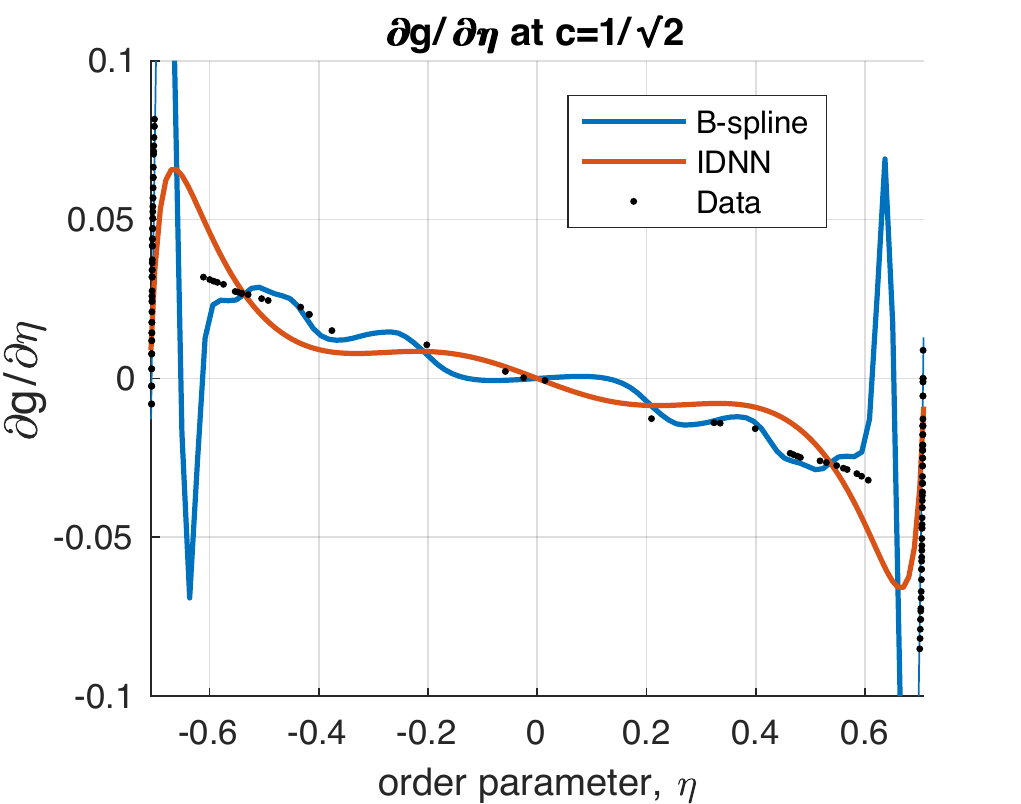}
\caption{Comparison of $\partial g/\partial\eta$ at $c = 1/\sqrt{2}$ for the IDNN and B-spline fits, showing the regions where the sign of $\partial g/\partial\eta$ goes from positive to negative and antiphase separation occurs. The $\partial g/\partial\eta$ data are also plotted as points.}
\label{fig:dg_deta}
\end{figure}

% \begin{figure}[tb]
%         \centering
% \begin{minipage}[t]{0.5\textwidth}
%         \centering
% 	\includegraphics[width=0.99\textwidth]{figures/DNN_2nd_der-2.pdf}
% 	\subcaption{DNN representation}
% \end{minipage}%
% \begin{minipage}[t]{0.5\textwidth}
%         \centering
% 	\includegraphics[width=0.99\textwidth]{figures/spline_2nd_der-2.pdf}
% 	\subcaption{B-spline representation}
% \end{minipage}
% \caption{Comparison of the second derivative of the free energy density with respect to the order parameter $\eta$ at $c = 1/\sqrt{2}$, showing the regions of convexity and non-convexity of the the DNN and B-spline representations of the free energy . The numerically differentiated data are also plotted as points.}
% \label{fig:2nd_der}
% \end{figure}

% \begin{figure}[tb]
%         \centering
% \begin{minipage}[t]{0.5\textwidth}
%         \centering
% 	\includegraphics[width=0.9\textwidth]{figures/bspline_0.png}
% 	\subcaption{Time step 0 ($t=0$)}
% \end{minipage}%
% \begin{minipage}[t]{0.5\textwidth}
%         \centering
% 	\includegraphics[width=0.9\textwidth]{figures/bspline_135.png}
% 	\subcaption{Time step 135 ($t=355$)}
% \end{minipage}
% \caption{The lack of nonconvexity in the B-spline representation of the free energy at $\eta=0$, $c=1/\sqrt{2}$ prevents the appearance of antiphase domains. The equilibrium solution in this case shows a uniform order parameter field of $\eta=0$.}
% \label{fig:spine_pf}
% \end{figure}

The behavior of the dynamics related to antiphase domains is dictated primarily by the sign of $\partial g/\partial\eta$. These values are plotted in Figure \ref{fig:dg_deta}, along with data points within $c = 1/\sqrt{2} \pm 1.1\times10^{-4}$ for comparison. Antiphase segregation occurs in a domain where the sign of $\partial g/\partial\eta$ changes from positive to negative for increasing $\eta$. In order for antiphase domains to occur in the phase field simulations, the initial conditions, which as in Section \ref{sec:pfresults} consist of random perturbations about $\eta = 0$ in the range (-0.01,0.01), must lie on both sides of the point where the sign of $\partial g/\partial\eta$ changes from positive to negative. While this condition holds true for the DNN representation, it does not hold for the B-spline representation. Due to oscillations in the B-spline representation, $\partial g/\partial\eta$ changes from negative to positive, rather than positive to negative, within the range of initial order parameter values. This results in a phase field solution that does not produce antiphase regions, but instead reaches an equilibrium with uniform values of $\eta=0$ and $c=1/\sqrt{2}$. It is possible, however, that other methods for knot selection or additional constraints might produce a better fit with B-splines.

The B-spline representation is more computationally efficient than the DNN. The phase field code evaluates the free energy and all first and second derivatives at each quadrature point. Using the de Boor algorithm for B-spline evaluation \cite{deboor1972}, the FLOP count for each function evaluation is about 1700. The highest order term in the count is $\sim165p^2$, with polynomial order $p$. Significantly, the FLOP count for the B-spline evaluation does not depend on the size of the knot vector. The DNN representation, on the other hand, requires about 4500 FLOPs per evaluation of the DNN and its derivatives, using a naive implementation. The highest order term of the count is $\sim18m^2n$, where $m$ is the neurons per layer and $n$ is the number of layers. However, an improved evaluation algorithm could potentially reduce the FLOP count. The use of accelerators could also speed up the function evaluations, without reducing the total FLOP count.

While the FLOP count for the free energy evaluation of the DNN is more than 2.5 times that of the B-spline, this affects only the computation time for the assembly of the residual vector and tangent matrix. The wall time for the matrix-vector solution will be equivalent for the two methods, assuming both fitted functions are similar enough to give comparable condition numbers. For the DNN and B-spline fits presented here, the average computation time per time step over the first 5 time steps using B-splines was 22.8 s, while it was 24.7 s using the DNN. The total computation time using the DNN was, then, only about 1.1 times that of the computation using the B-spline to represent the free energy. For larger problems, the matrix size increases and solver time comes to dominate the average computation time. In this limit the wall times will converge.

\section{Conclusions}
\label{sec:conclusions}
This communication adds to our nascent, but growing body of work in machine learning and artificial intelligence targeting higher fidelity models of materials physics \cite{natarajan2018,Teichert2018a}. We have explored machine learning as an approach to bridging scales, by focusing on the representation of complexity emerging from fine scale physics. Here, the fine scales come from atomic and statistical mechanics descriptions, which can parameterize thermodynamic functions, such as free energy densities with high fidelity. At the core of this work is the idea of an analytically integrable Deep Neural Network (IDNN) to represent such functions. The IDNN is of particular use in the context of mathematically representing the free energy density of a material, where only the derivative data is originally known and the trained function must be integrated to recover the free energy. It is highly relevant to multidimensional systems, where multiple sets of partial derivative data must be trained against simultaneously under the constraint that all trained functions are the partial derivatives of one common function.

Using as a prototypical case a binary alloy, an IDNN was trained to two sets of chemical potential data, which were found using first principles calculations. Since these datasets are derivatives of the same free energy density, the analytic integrability of the IDNN, which exactly preserves consistency of representation, assumes importance. Symmetry with respect to the order parameter was embedded in the IDNN. We anticipate an expansion of these approaches to embedding fundamental aspects of the physics into machine learning models as this field develops. 

Phase field simulations with the analytically integrated DNN representing the free energy recovered the proper physics of the system, showing the creation and subsequent coarsening of antiphase domains in the material. This example demonstrates the ability of the IDNN to capture the relevant physics of a material system. Interestingly, the IDNN was able to represent the physics of the system more faithfully than a B-spline representation, even in the current, relatively simple case, of a two-dimensional input. We note that in earlier work, we have demonstrated that B-spline representations themselves are superior to the more traditional Redlich-Kister polynomials at resolving rapidly varying thermodynamic functions \cite{Teichert2017}. The present work is a continuation of that thread, and establishes that, even for  only two-dimensional functions, the B-spline approach also may be inadequate. It is natural to expect that high-dimensional chemical potentials and free energy densities will present greater challenges.

Future work will consider systems of greater complexity, where the free energy density has such higher-dimensional dependence on variables that number $\sim \mathcal{O}(10)$. In this regime, high fidelity representations of the free energy density will be crucial to reproducing the physics through scale bridging, and the uniform approximation property of DNNs, inherited by the IDNNs, will deliver greater advantages. 

\section{Acknowledgements}
\label{sec:acknowledgements}
We gratefully acknowledge the support of Toyota Research Institute, Award \#849910, ``Computational framework for data-driven, predictive, multi-scale and multi-physics modeling of battery materials". This work has also been supported in part by National Science Foundation DMREF grant \#1729166, ``Integrated Framework for Design of Alloy-Oxide Structures''. Simulations in this work were performed using the Extreme Science and Engineering Discovery Environment (XSEDE) Comet at the San Diego Supercomputer Center through allocations TG-MSS160003 and TG-DMR180072. XSEDE is supported by National Science Foundation grant number ACI-1548562. Computing resources also were provided in part by the NSF via grant 1531752 MRI: Acquisition of Conflux, A Novel Platform for Data-Driven Computational Physics (Tech. Monitor: Ed Walker).

\bibliographystyle{unsrt}
\bibliography{references}

\end{document}